\documentclass[english,aps,manuscript]{revtex4}
\usepackage[T1]{fontenc}
\usepackage[latin9]{inputenc}
\usepackage{amstext}
\usepackage{amssymb}
\usepackage{graphicx}
\usepackage{esint}

\makeatletter

\DeclareRobustCommand{\greektext}{%
  \fontencoding{LGR}\selectfont\def\encodingdefault{LGR}}
\DeclareRobustCommand{\textgreek}[1]{\leavevmode{\greektext #1}}
\DeclareFontEncoding{LGR}{}{}
\DeclareTextSymbol{\~}{LGR}{126}
\providecommand{\tabularnewline}{\\}

\@ifundefined{textcolor}{}
{%
 \definecolor{BLACK}{gray}{0}
 \definecolor{WHITE}{gray}{1}
 \definecolor{RED}{rgb}{1,0,0}
 \definecolor{GREEN}{rgb}{0,1,0}
 \definecolor{BLUE}{rgb}{0,0,1}
 \definecolor{CYAN}{cmyk}{1,0,0,0}
 \definecolor{MAGENTA}{cmyk}{0,1,0,0}
 \definecolor{YELLOW}{cmyk}{0,0,1,0}
}

\makeatother

\usepackage{babel}
\begin{document}

\title{Complex Saddle Points and Disorder Lines in QCD at finite temperature
and density}

\author{Hiromichi Nishimura}

\address{Faculty of Physics, University of Bielefeld, D-33615 Bielefeld, Germany}

\author{Michael C. Ogilvie and Kamal Pangeni}

\address{Department of Physics, Washington University, St. Louis, MO 63130
USA}

\date{11/17/14}
\begin{abstract}
The properties and consequences of complex saddle points are explored
in phenomenological models of QCD at non-zero temperature and density.
Such saddle points are a consequence of the sign problem, and should
be considered in both theoretical calculations and lattice simulations.
Although saddle points in finite-density QCD are typically in the
complex plane, they are constrained by a symmetry that simplifies
analysis. We model the effective potential for Polyakov loops using
two different potential terms for confinement effects, and consider
three different cases for quarks: very heavy quarks, massless quarks
without modeling of chiral symmetry breaking effects, and light quarks
with both deconfinement and chiral symmetry restoration effects included
in a pair of PNJL models. In all cases, we find that a single dominant
complex saddle point is required for a consistent description of the
model. This saddle point is generally not far from the real axis;
the most easily noticed effect is a difference between the Polyakov
loop expectation values $\left\langle {\rm Tr}_{F}P\right\rangle $
and $\left\langle {\rm Tr}_{F}P^{\dagger}\right\rangle $, and that
is confined to small region in the $\mu-T$ plane. In all but one
case, a disorder line is found in the region of critical and/or crossover
behavior. The disorder line marks the boundary between exponential
decay and sinusoidally modulated exponential decay of correlation
functions. Disorder line effects are potentially observable in both
simulation and experiment. Precision simulations of QCD in the $\mu-T$
plane have the potential to clearly discriminate between different
models of confinement.
\end{abstract}
\maketitle

\section{Introduction}

The phase structure of QCD at finite density and temperature is of
fundamental importance, and can be studied experimentally, theoretically
and via lattice simulations. Nevertheless, progress has been slow,
in part because of the sign problem, which afflicts both phenomenological
models and lattice simulations. The sign problem is found in many
area of physics \cite{deForcrand:2010ys,Gupta:2011ma,Aarts:2013bla}.
In QCD, the quark contribution to the partition function, given as
a functional determinant dependent on the gauge field, is complex
for typical gauge field configurations when the quark chemical potential
$\mu$ is non-zero. It is natural to consider the analytically continuation
of the gauge field into the complex plane. Some progress has been
made in incorporating this idea into lattice simulations \cite{Cristoforetti:2012uv,Fujii:2013sra,Cristoforetti:2013qaa,Cristoforetti:2013wha,Cristoforetti:2014gsa,Cristoforetti:2014waa,Sexty:2014dxa}.
Here we show that the consideration of complex saddle points provides
a conceptually cohesive phenomenological model of QCD at finite $T$
and $\mu$. Our results can provide guidance for lattice simulations
by indicating the behavior of the dominant field configuration, within
a phenomenological framework. We will show that certain features of
the saddle point appear to be independent of the choice of a particular
phenomenological model. Moreover, we will identify a new property
of QCD at finite density, the occurence of a disorder line, that may
have observable consequences in experiment and/or lattice simulation.
Some feature associated with the disorder line differentiate strongly
between different phenomenological models, and may thus have an impact
on our understanding of confinement.

The remainder of this paper is organized as follows. In section II,
we provide a simple example based on the the $U(1)$ group that indicates
the need for complex saddle points. Section III reviews the formalism
first developed in our previous work \cite{Nishimura:2014rxa}. We
pay particular attention to the existence and consequences of an antilinear
symmetry $\mathcal{CK}$ in finite density field theories, where $\mathcal{C}$
is charge conjugation and $\mathcal{K}$ is complex conjugation; in
some sense this symmetry replaces charge conjugation symmetry when
$\mu\ne0$. The following section, section IV, describes the different
phenomenological models we study using an effective potential for
the Polyakov loop $P$ and chiral condensate $\bar{\psi}\psi$. We
do not consider other possible condensates in this work, such as the
color superconducting condensate, deferring this to later work. A total
of six different models are considered. We use two different models
for the confining part of the effective potential, Model A and Model
B, taken from \cite{Meisinger:2001cq}. We consider three cases of
quarks, always with two flavors: heavy quarks, massless quarks with
no chiral dynamics and a full treatment of light quarks, with chiral
dynamics included via a bosonized four-fermion interaction. Our most
realisitic models are therefore of Polyakov-Nambu-Jona Lasinio (PNJL)
type, with the major new feature the consideration of complex saddle
points of the effective potential. Section V, VI and VII describe
in detail the results for the three different cases of quarks. A final
section offers conclusions.

\section{Simple U(1) example}

As an illustration of the role of analytic continuation in field space
for models with non-zero chemical potential, we consider a single-site
model, where a particle propagates in a closed loop in Euclidean time,
always returning to the same lattice site. The model has a hopping
parameter $J$, a dimensionless chemical potential $\mu$ and a $U(1)$
background field $\theta$ \cite{Aarts:2008rr}. The partition function
is
\begin{equation}
Z=\int\frac{d\theta}{2\pi}\, e^{S}
\end{equation}
 where
\begin{equation}
S=J\left[e^{\mu+i\theta}+e^{-\mu-i\theta}\right].
\end{equation}
The action $S$ is complex, so $Z$ has a sign problem. It is easy
to find $Z$ exactly by a strong-coupling expansion in $J$:
\begin{equation}
Z=\sum_{n=0}^{\infty}\frac{J^{2n}}{\left(n!\right)^{2}}=I_{0}\left(J\right),
\end{equation}
where $ $$I_{0}\left(J\right)$ is the modified Bessel function of
order $0$. Similar results can be obtained for expectation values
such as $\left\langle e^{i\theta}\right\rangle $, which are zero-dimensional
analogs of Polyakov loops. It is instructive to consider $Z$ as a
contour integral in the variable $z=\exp\left(i\theta\right)$:
\begin{equation}
Z=\int_{\left|z\right|=1}\frac{dz}{2\pi iz}\,\exp\left[Jze^{\mu}+Jz^{-1}e^{-\mu}\right].
\end{equation}
We ask if the contour $\left|z\right|=1$ can be deformed to a contour
$C$ along which $S$ is real. The contour $C$ is given by the circle
$\left|z\right|=e^{-\mu}$. Making a change of variable $\theta\rightarrow\theta+i\mu$,
we recover exact results such as
\begin{eqnarray}
Z & = & I_{0}\left(J\right)\nonumber \\
\left\langle e^{i\theta}\right\rangle  & = & e^{-\mu}\frac{I_{1}\left(J\right)}{I_{0}\left(J\right)}\nonumber \\
\left\langle e^{-i\theta}\right\rangle  & = & e^{+\mu}\frac{I_{1}\left(J\right)}{I_{0}\left(J\right)}.\label{eq:toy_exact}
\end{eqnarray}
We apply a saddle-point method to the original integral, looking for
the saddle-point in the complex plane. The saddle point satisfies
\begin{equation}
e^{\mu}-e^{-\mu}/z^{2}=0
\end{equation}
 so the saddle is at $i\theta=-\mu$. Returning to the original notation,
we approximate $Z$ by
\begin{equation}
Z\approx\int\frac{d\theta}{2\pi}\exp\left[2J-\frac{1}{2}2J\theta^{2}\right]=\frac{e^{2J}}{\sqrt{4\pi J}}
\end{equation}
which is the leading-order asymptotic behavior of $I_{0}\left(2J\right)$.
A similar evaluation for the expectation values yields
\begin{eqnarray}
\left\langle e^{i\theta}\right\rangle  & \simeq & e^{-\mu}\nonumber \\
\left\langle e^{-i\theta}\right\rangle  & \simeq & e^{+\mu}.\label{eq:toy_expec}
\end{eqnarray}
If we had used $Re(S)=J\left(e^{\mu}+e^{-\mu}\right)\cos\theta$ as
a starting point for a steepest descents calculation, the result for
$Z$ would have been
\begin{equation}
\frac{e^{J\left(e^{\mu}+e^{-\mu}\right)}}{\sqrt{2\pi J\left(e^{\mu}+e^{-\mu}\right)}}
\end{equation}
 which does not represent the correct asymptotic behavior.

It is important to emphasize that neither a deformation of the contour
into the complex plane nor the use of complex saddle points is required
in an exact evaluation of $Z$ and related quantities. However, many
methods, from perturbation theory to importance sampling in lattice
simulations, rely implicitly or explicitly on the existence of appropriate
saddle points. In this simple $U(1)$ model, the use of complex saddle
points naturally allows the expected values of the Polyakov loops
for particle and antiparticles to be different: $\left\langle e^{i\theta}\right\rangle \ne\left\langle e^{-i\theta}\right\rangle $.
In an exact calculation using a real contour for $\theta$, this result
must be recovered from rapid fluctuations in the integration. A saddle
point approximation incorrectly using $Re(S)$ for the location of
saddle points would have obtained $\left\langle e^{i\theta}\right\rangle =\left\langle e^{-i\theta}\right\rangle $
at leading order.

\section{Formalism for SU(N) gauge theories at finite density}

We now consider an $SU(N)$ gauge theory coupled to fermions in the
fundamental representation. It is well-known that the Euclidean Dirac
operator has complex eigenvalues when a non-zero chemical potential
is introduced \cite{deForcrand:2010ys}. This can be understood as
an explicit breaking of charge conjugation symmetry $\mathcal{C}$.
The log of the fermion determinant, $\log\det\left(\mu,A\right)$,
which is a function of the quark chemical potential $\mu$ and the
gauge field $A$, can be formally expanded as a sum over Wilson loops
with real coefficients. For a gauge theory at finite temperature,
the sum includes Wilson loops that wind non-trivially around the Euclidean
timelike direction; Polyakov loops, also known as Wilson lines, are
examples of such loops. At $\mu=0$, every Wilson loop ${\rm Tr}_{F}W$
appearing in the expression for the fermion determinant is combined
with its conjugate ${\rm Tr}_{F}W^{\dagger}$ to give a real contribution
to path integral weighting. More formally, charge conjugation acts
on matrix-valued Hermitian gauge fields as

\begin{equation}
\mathcal{C}:\, A_{\mu}\rightarrow-A_{\mu}^{t}
\end{equation}
where the overall minus sign is familiar from QED, and the transpose
interchanges particle and antiparticle, \emph{e.g.}, $W^{+}$ and
$W^{-}$ in $SU(2).$ This transformation law in turn implies that
$\mathcal{C}$ exchanges ${\rm Tr}_{F}W$ and ${\rm Tr}_{F}W^{\dagger}$
so unbroken charge symmetry implies a real fermion determinant. When
$\mu\ne0$, Wilson loops with non-trivial winding number $n$ in the
$x_{4}$ direction receive a weight $e^{n\beta\mu}$ while the conjugate
loop is weighted by $e^{-n\beta\mu}$ and invariance under $\mathcal{C}$
is explicitly broken. However, there is a related antilinear symmetry
which is unbroken: ${\rm Tr}_{F}W$ transforms into itself under the
combined action of $\mathcal{CK}$, where $\mathcal{K}$ is the fundamental
antilinear operation of complex conjugation. Thus the theory is invariant
under $\mathcal{CK}$ even in the case $\mu\ne0$. This symmetry is
an example of a generalized $\mathcal{PT}$ (parity-time) symmetry
transformation \cite{Bender:1998ke,Meisinger:2012va}; theories with
such symmetries form special class among theories with sign problems.
For fermions, $\mathcal{CK}$ symmetry implies the well-known relation
$\det\left(-\mu,A_{\mu}\right)=\det\left(\mu,A_{\mu}\right)^{*}$
for Hermitian $A_{\mu}$, a relation which is often derived using
a $\gamma_{5}$ transformation of the Dirac operator. The advantage
of using $\mathcal{CK}$ is that it is more general, leading to more
insight into the sign problem and applying to bosons as well as to
fermons. For example, it is easy to see that our simple zero-dimensional
$U(1)$ model in the preceding section is invariant under the combined
action of $\mathcal{K}:i\rightarrow-i$ and $\mathcal{C}:\theta\rightarrow-\theta$. 

For phenomenological models, the existence of $\mathcal{CK}$ symmetry
leads naturally to the consideration of complex but $\mathcal{CK}$-symmetric
saddle points. $\mathcal{CK}$ symmetry will map any saddle-point
configuration $A_{\mu}^{(1)}$ into another saddle point given by
$A_{\mu}^{(2)}=-A_{\mu}^{(1)\dagger}$ with a corresponding connection
between the actions of the two configurations: $S^{(2)}=S^{(1)*}$.
However, some field configurations are themselves $\mathcal{CK}$-symmetric
in that $-A_{\mu}^{\dagger}$ is equivalent to $A_{\mu}$ under a
gauge transformation. If a saddle point is $\mathcal{CK}$ symmetric,
then its action and effective potential are necessarily real. A quick
direct proof can be given: For such a field configuration, it is easy
to prove that every Wilson loop is real and thus $\det\left(\mu,A_{\mu}\right)$
is real and positive for a $\mathcal{CK}$-symmetric field configuration.
If a single $\mathcal{CK}$-symmetric saddle point dominates the effective
potential, then the sign problem is solved, at least for a particular
phenomenological model. Such $\mathcal{CK}$-symmetric saddle points
have been seen before in finite density calculations \cite{Hands:2010vw,Hands:2010zp,Hollowood:2011ep,Hollowood:2012nr}. 

Let us consider the Polyakov loop $P$, a special kind of Wilson loop,
associated with some particular field configuration that is $\mathcal{CK}$-symmetric.
We can transform to Polyakov gauge where $A_{4}$ is diagonal and
time-independent, and work with the eigenvalues $\theta_{j}$ defined
by
\begin{equation}
P\left(\vec{x}\right)=\mbox{diag}\left[e^{i\theta_{1}\left(\vec{x}\right)},\cdots,\, e^{i\theta_{N}\left(\vec{x}\right)}\right]
\end{equation}
where the $\theta_{j}$'s are complex but satisfy $\sum_{j}\theta_{j}=0$.
Because we are primarily interested in constant saddle points, we
suppress the spatial dependence hereafter. Invariance under $\mathcal{CK}$
means that the set $\left\{ -\theta_{j}^{*}\right\} $ is equivalent
to the $\left\{ \theta_{j}\right\} $ although the eigenvalues themselves
may permute. In the case of $SU(3)$, we may write this set uniquely
as
\begin{equation}
\left\{ \theta-i\psi,-\theta-i\psi,2i\psi\right\} .\label{eq:EVs_PolyakovLoop_SU3}
\end{equation}
This parametrizes the set of $\mathcal{CK}$-symmetric $SU(3)$ Polyakov
loops. Notice that both
\begin{equation}
{\rm Tr}_{F}P=e^{\psi}2\cos\theta+e^{-2\psi}
\end{equation}
 and
\begin{equation}
{\rm Tr}_{F}P^{\dagger}=e^{-\psi}2\cos\theta+e^{2\psi}
\end{equation}
are real, but they are equal only if $\psi=0$. In the usual interpretation
of the Polyakov loop expectation value, this implies that the free
energy change associated with the insertion of a fermion is different
from the free energy change associated with its antiparticle. It is
easy to check that the trace of all powers of $P$ or $P^{\dagger}$
are all real, and thus all group characters are real as well. This
parametrization represents a generalization of the Polyakov loop parametrization
used in the application of mean-field methods to confinement, $e.g.$,
in PNJL models \cite{Fukushima:2003fw} or in gauge theories with
double-trace deformations \cite{Myers:2007vc,Unsal:2007vu}. This
parametrization can be generalized to include finite density models
for arbitrary $N$.

The existence of complex $\mathcal{CK}$-symmetric saddle points provides
a fundamental approach to non-Abelian gauge theories that is similar
to the heuristic introduction of color chemical potentials, and naturally
ensures the system has zero color charge, \emph{i.e.}, all three charges
contribute equally \cite{Buballa:2005bv}. In the case of $SU(3)$,
extremization of the thermodynamic potential with respect to $\theta$
leads to the requirement $\left\langle n_{r}\right\rangle -\left\langle n_{g}\right\rangle =0$
where $\left\langle n_{r}\right\rangle $ is red color density, including
the contribution of gluons. Similarly, extremization of the thermodynamic
potential with respect to $\psi$ leads $\left\langle n_{r}\right\rangle +\left\langle n_{g}\right\rangle -2\left\langle n_{b}\right\rangle =0$.
Taken together, these two relations imply that $\left\langle n_{r}\right\rangle =\left\langle n_{g}\right\rangle =\left\langle n_{b}\right\rangle $.

We demand that any saddle point solution be stable to constant, real
changes in the Polyakov loop eigenvalues, corresponding for $SU(3)$
to constant real changes in $A_{4}^{3}$ and $A_{4}^{8}$. Consider
the $\left(N-1\right)\times\left(N-1\right)$ matrix $M_{ab}$, defined
in Polyakov gauge as 
\begin{equation}
M_{ab}\equiv g^{2}\frac{\partial^{2}V_{eff}}{\partial A_{4}^{a}\partial A_{4}^{b}}.
\end{equation}
At very high temperatures and densities, the eigenvalues of this mass
matrix give the usual Debye screening masses. The stability criterion
is that the eigenvalues of $M$ must have positive real parts. At
$\mathcal{CK}$-symmetric saddle points, the eigenvalues will be either
real or part of a complex conjugate pair. In the case of $SU(3),$
the matrix $M$ may also be written in terms of derivatives with respect
to $\theta$ and $\psi$ as
\begin{equation}
M=\frac{g^{2}}{T^{2}}\left(\begin{array}{cc}
\frac{1}{4}\frac{\partial^{2}V_{eff}}{\partial\theta^{2}} & \frac{i}{4\sqrt{3}}\frac{\partial^{2}V_{eff}}{\partial\theta\partial\psi}\\
\frac{i}{4\sqrt{3}}\frac{\partial^{2}V_{eff}}{\partial\theta\partial\psi} & \frac{-1}{12}\frac{\partial^{2}V_{eff}}{\partial\psi^{2}}
\end{array}\right).\label{eq:M_mass_matrix}
\end{equation}
This stability criterion generalizes the stability criterion used
previously for color chemical potentials, which was $\partial^{2}V_{eff}/\partial\psi^{2}<0$.
Crucially, the mass matrix $M_{ab}$ is invariant under $M^{*}=\sigma_{3}M\sigma_{3}$,
which is itself a generalized $\mathcal{PT}$ (parity-time) symmetry
transformation \cite{Bender:1998ke,Meisinger:2012va}. It is easy
to see that this relation implies that $M_{ab}$ has either two real
eigenvalues or a complex eigenvalue pair. In either case, the real
part of the eigenvalues must be positive for stability. In the case
where there are two real eigenvalues, we will denote by $\kappa_{1}$
and $\kappa_{2}$ the two positive numbers such that $\kappa_{1}^{2}$
and $\kappa_{2}^{2}$ are the eigenvalues of the mass matrix $M_{ab}$.
If $M_{ab}$ has two complex eigenvalues, we define two positive real
numbers $\kappa_{R}$ and $\kappa_{I}$ such that $\left(\kappa_{R}\pm i\kappa_{I}\right)^{2}$
are the conjugate eigenvalues of $M_{ab}.$ The border separating
the region $\kappa_{I}\ne0$ from the region $\kappa_{I}=0$ is known
as the disorder line \cite{StephensonJPhysRevB.1.4405,StephensonJCanJPdoi:10.1139/p70-217,Selke1988213}.
In this case, it separates the region where the color density correlation
function decays exponentially in the usual way from the region where
a sinusoidal modulation is imposed on that decay.

We illustrate the working of $\mathcal{CK}$ symmetry using the well-known
one-loop expressions for the effective potential of particles moving
in a constant background Polyakov loop. The one-loop contribution
to the effective potential of $N_{f}$ flavors of fundamental fermions
moving in a background gauge field $A$ is given by
\begin{equation}
\beta\mathcal{V}V_{eff}^{f}=-N_{f}\log\left[\det\left(\mu,A\right)\right]
\end{equation}
where $\det$ again represents the functional determinant of the Dirac
operator and $\beta\mathcal{V}$ is the volume of spacetime. A compact
expression for the effective potential of massless fermions when the
eigenvalues of $P$ are complex was derived using zeta function methods
in \cite{KorthalsAltes:1999cp}. The finite temperature contribution
to the effective potential from quarks is given by 
\begin{equation}
V_{f}^{T}\left(P\right)=-2TN_{f}\int\frac{d^{3}k}{\left(2\pi\right)^{3}}{\rm Tr}_{F}\left[\log\left(1+e^{\beta\mu-\beta\omega_{k}}P\right)+\log\left(1+e^{-\beta\mu-\beta\omega_{k}}P^{\dagger}\right)\right]\label{eq:Vf_def_1loop}
\end{equation}
where $\omega_{k}=+\sqrt{k^{2}+m^{2}}$ with $m$ the fermion mass.
We have evaluated $V_{f}(P)$ analytically for the case of massless
quarks \cite{Nishimura:2014rxa}. The result for quarks in a $\mathcal{CK}$-symmetric
$SU(3)$ background Polyakov loop is
\begin{equation}
V_{f}^{T}(\theta,\psi,\text{T},\mu)=N_{f}\left(v_{f}\left(\theta-i\psi-\frac{i\mu}{T}\right)+v_{f}\left(-\theta-i\psi-\frac{i\mu}{T}\right)+v_{f}\left(2i\psi-\frac{i\mu}{T}\right)\right)
\end{equation}
where
\begin{equation}
v_{f}(\theta)=-\frac{4T^{4}}{\pi^{2}}\left(\frac{\theta^{4}}{48}-\frac{\pi^{2}\theta^{2}}{24}+\frac{7\pi^{4}}{720}\right).
\end{equation}
Explicitly, this is 
\begin{eqnarray}
V_{f}^{T}(\theta,\psi,\text{T},\mu) & = & -\frac{\mu^{4}}{2\pi^{2}}+T^{2}\left(-\mu^{2}+\frac{2\theta^{2}\mu^{2}}{\pi^{2}}-\frac{6\mu^{2}\psi^{2}}{\pi^{2}}\right)+\frac{4T^{3}\left(\theta^{2}\mu\psi+\mu\psi^{3}\right)}{\pi^{2}}\label{eq:Vf_massless}\nonumber\\
 &  & +\frac{T^{4}\left(-7\pi^{4}+20\pi^{2}\theta^{2}-10\theta^{4}-60\pi^{2}\psi^{2}+60\theta^{2}\psi^{2}-90\psi^{4}\right)}{30\pi^{2}}.
\end{eqnarray}
for two flavors of massless quarks. This is manifestly real. Because
we are interested in the analytic continuation of Polyakov loop eigenvalues
into the complex plane, we need expressions for the gauge bosons as
well as for fermions. In our previous work, we have shown that for
$SU(3)$

\begin{equation}
V_{g}(P)=\frac{T^{4}\left(135\left(\theta^{2}-3\psi^{2}\right)^{2}+180\pi^{2}\left(\theta^{2}-3\psi^{2}\right)+60\pi\theta\left(27\psi^{2}-5\theta^{2}\right)-16\pi^{4}\right)}{90\pi^{2}}\label{eq:Vg}
\end{equation}
which is also manifestly real. Note that the valid range of $\theta$
is $\left(0,\pi\right)$ due to the appearance of $2\theta$ as an
eigenvalue in the adjoint representation. The one-loop effective potential
is simply the sum of $V_{g}(\theta)$ and $V_{f}(\theta)$. As is
the case when $\mu=0$, the dominant saddle point remains at $\theta=0$
when $\mu\ne0$: the one-loop effective potential incorrectly predicts
that QCD is always in the extreme deconfined phase with ${\rm Tr}_{F}P={\rm Tr}_{F}P^{\dagger}=3$
because there is no confinement mechanism included.

\section{Models}

We now consider a class of phenomenological models that combines the
one-loop result with the effects of confinement for the case of $SU(3)$
gauge bosons and two flavors of quarks at finite temperature and density.
The model is described by an effective potential which is the sum
of three terms: 
\begin{equation}
V_{eff}(P)=V_{g}(P)+V_{f}(P)+V_{d}(P).
\end{equation}
The potential term $V_{g}(P)$ is the one-loop effective potential
for gluons given by eqn.\ (\ref{eq:Vg}). The potential term $V_{f}\left(P\right)$
contains all quark effects, including the one-loop expression defined
above in eqn.\ (\ref{eq:Vf_def_1loop}). The potential term $V_{d}\left(P\right)$
represents confinement effects. We will consider three different forms
for $V_{f}\left(P\right)$ and two different forms for $V_{d}\left(P\right)$
for a total of six different models. The formulas and parameters we
use for these models are summarized in Tables \ref{tab:confinement_parameters}
and \ref{tab:chiral_parameters}.

The potential term $V_{d}(P)$ acts to favor the confined phase at
low temperature and density \cite{Meisinger:2001cq,Myers:2007vc,Unsal:2008ch,Ogilvie:2012is}.
There are two different points of view that can be taken on this potential.
In one view, $V_{d}(P)$ represents a deformation added to the original
model, and hence the subscript $d$. In typical applications, the
temperature $T$ is taken to be large such that perturbation theory
is reliable in the chromoelectric sector because the running coupling
$g^{2}\left(T\right)$ is small. The deformation term is taken to
respect center symmetry and is used to move between the confined and
deconfined phases in a controlled way. The gauge contribution $V_{g}(P)$
favors the deconfined phase, and in the pure gauge theory ($N_{f}=0$)
the deconfinement transition arises out of the competition between
$V_{g}(P)$ and $V_{d}(P)$. The confined phase arising in models
of this type is known to be analytically connected to the usual low-temperature
confined phase of $SU(3)$ gauge theory \cite{Myers:2007vc}. This
point of view emphasizes analytic control at the price of deforming
the original gauge theory by the addition of $V_{d}(P)$. In the second
point of view, $V_{d}$ is phenomenological in nature and models the
unknown confining dynamics of the pure gauge theory. The parameters
of $V_{d}(P)$ are set to reproduce the deconfinement temperature
of the pure gauge theory, known from lattice simulations to occur
at $T_{d}\approx270\,\mbox{MeV}$. 

\begin{figure}
\includegraphics[width=5in]{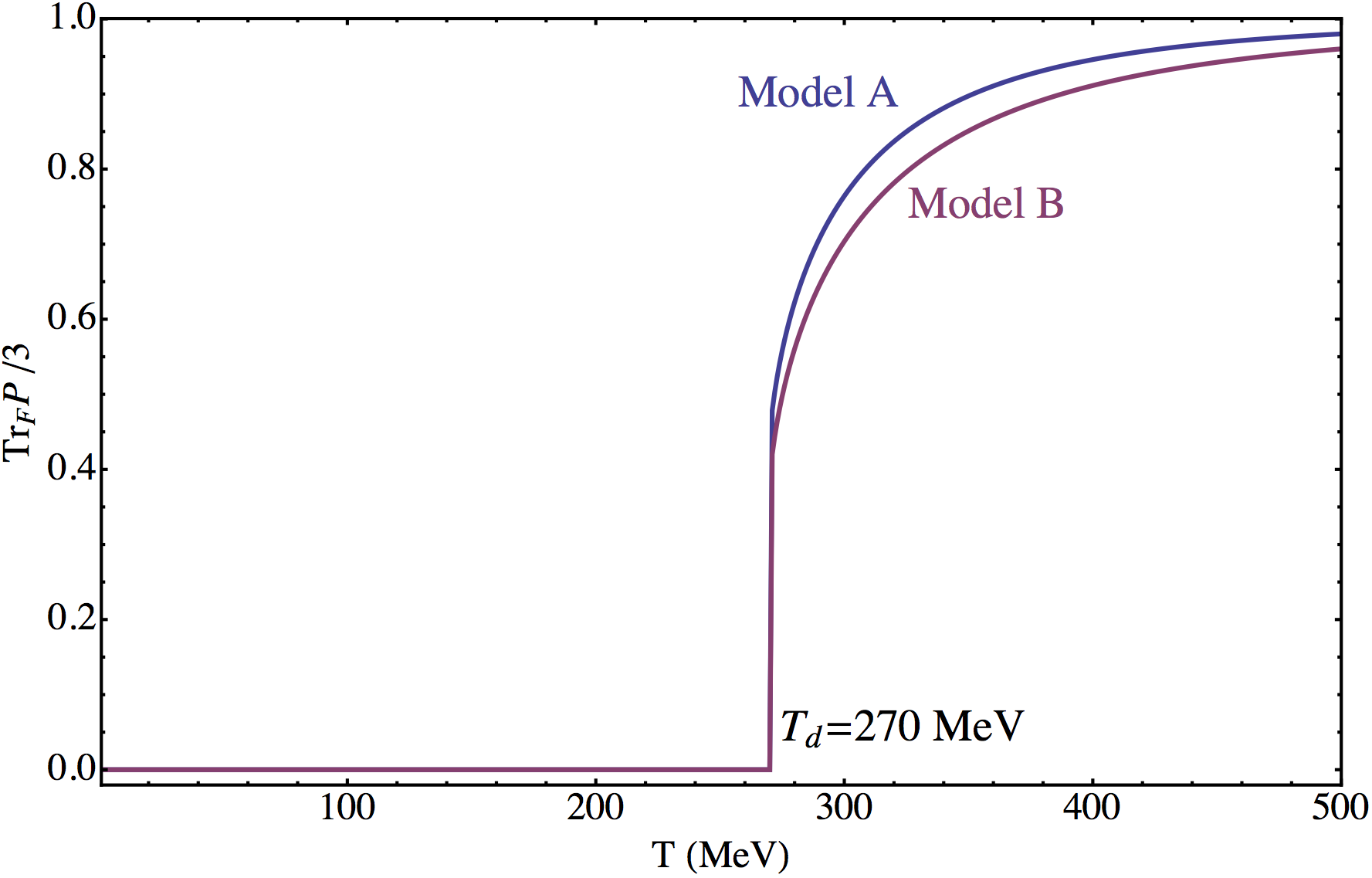}

\caption{\label{fig:PureGauge_AvsB}$\left\langle \mbox{Tr}_{F}P\right\rangle $
as a function of $T$ for pure $SU(3)$ with Model A and Model B for
confinement effects.}
\end{figure}

We will take the second point of view, using simple expressions for
$V_{d}(P)$ that reproduces much of the thermodynamic behavior seen
in lattice simulations of the pure gauge theory. The specific form
used are Model A and Model B of \cite{Meisinger:2001cq}. In Model
A, $V_{d}(P)$ can be written as 

\begin{equation}
V_{d}^{A}\left(P\right)=\sum_{j,k=1}^{N}(1-\frac{1}{N}\delta_{jk})\frac{M_{A}^{2}}{2\beta^{2}}B_{2}\left(\frac{\Delta\theta_{jk}}{2\pi}\right)
\end{equation}
where $\Delta\theta_{jk}=\left|\theta_{j}-\theta_{k}\right|$ are
the adjoint Polyakov loop eigenvalues and $B_{2}$ is the second Bernoulli
polynomial. This expression gives a simple quartic polynomial in the
Polyakov loop eigenvalues for $V_{g}\left(P\right)+V_{d}^{A}\left(P\right)$
and thus can be thought of as a form of Landau-Ginsburg potential
for the Polyakov loop eigenvalues. For the $SU(3)$ parametrization
used here, $V_{d}^{A}(P)$ takes the simple form
\begin{equation}
V_{d}^{A}\left(P\right)=\frac{M_{A}^{2}T^{2}\left((2\pi-3\theta)^{2}-27\psi^{2}\right)}{6\pi^{2}}.\label{eq:VA_d}
\end{equation}
The parameter $M_{A}$ controls the location of the deconfinement
transition in the pure gauge theory, and is set to $596\,\mbox{MeV}$.
At low temperatures, this term dominates the pure gauge theory effective
potential. The variable $\psi$ is zero, and $V_{d}\left(P\right)$
is minimized when $\theta=2\pi/3$. For this value of $\theta$, the
eigenvalues of $P$ are uniformly spaced around the unit circle, respecting
center symmetry, and $\mathrm{Tr}_{F}P=\mathrm{Tr}_{F}P^{\dagger}=0$.
As the temperature increases, $V_{g}\left(P\right)$ becomes relevant,
and gives rise to the deconfined phase where center symmetry is spontaneously
broken. The addition of light fundamental quarks via $V_{f}(P)$ explicitly
breaks center symmetry. For all nonzero temperatures, center symmetry
is broken and $\left\langle \mathrm{Tr}_{F}P\right\rangle \ne0$.
However, a remnant of the deconfinement transition remains in the
form of a rapid crossover from smaller value of $\mathrm{Tr}_{F}P$
to larger ones as $T$ and $\mu$ are varied. We also use Model B,
defined as
\begin{equation}
V_{d}^{B}(P)=-\frac{T}{R^{3}}\log\left[\prod_{j<k}\sin^{2}\left(\frac{\theta_{j}-\theta_{k}}{2}\right)\right].\label{eq:-1}
\end{equation}
This form for $V_{d}$ is motivated by Haar measure, representing
a determinantal term that tries to keep a space-time volume of order
$\beta R^{3}$ confined. For the $SU(3)$ parametrization, $V_{d}^{B}(P)$
takes the form
\begin{equation}
V_{d}^{B}(P)=-\frac{T}{R^{3}}\log\left[\frac{1}{4}\left\{ \cos\theta-\cosh\left(3\psi\right)\right\} ^{2}\sin^{2}\theta\right].\label{eq:VB_d}
\end{equation}
In order to reproduce the correct deconfinement temperature for the
pure gauge theory, $R$ must be set to $R=1.0028$ fm. We plot the
Polyakov loop for both Model A and Model B in Fig. \ref{fig:PureGauge_AvsB}.

Although $V_{d}^{A}$ and $V_{d}^{B}$ appear to be very different,
and are motivated in different ways, they are actually closely related.
The deformation potential $V_{d}^{A}$ can also be written as
\begin{equation}
V_{d}^{A}=\frac{M_{A}^{2}T^{2}}{2\pi^{2}}\sum_{n=1}^{\infty}\frac{1}{n^{2}}{\rm Tr}_{A}P^{n}
\end{equation}
while $V_{d}^{B}$ can be written as
\begin{equation}
V_{d}^{B}=\frac{T}{R^{3}}\sum_{n=1}^{\infty}\frac{1}{n}{\rm Tr}_{A}P^{n}.
\end{equation}
Using ${\rm Tr}_{A}P={\rm Tr}_{F}P^{n}{\rm Tr}_{F}P^{\dagger n}-1$,
it is easy to prove that minimizing either $V_{d}^{A}$ or $V_{d}^{B}$
yields a confining phase where ${\rm Tr}_{F}P^{n}=0$ for all $n\ne0\,\mathrm{mod}(N)$. 

We consider three different cases of quarks. The first is heavy quarks,
with a fixed mass of $2$ GeV. The form of $V_{f}\left(P\right)$
is precisely that of Eq. (\ref{eq:Vf_def_1loop}) with the fermion
mass set to a large value. In this model, the quarks are essentially
irrelevant for the deconfinement transition, which occurs at essentially
the same temperature as if no quarks were present at all. The effect
of spontaneous chiral symmetry breaking is not included, as it would
only contribute a small amount to the quark mass. This case is in
some sense the simplest, and perhaps would be the easiest for which
to obtain reliable simulation results. The second case considered
is massless quarks, where the fermion mass in Eq. (\ref{eq:Vf_def_1loop})
is set equal to zero. This case cannot be easily simulated using lattice
methods, because it ignores chiral symmetry breaking effects which
do occur in lattice simulations. It is thus useful only for sufficiently
large values of $T$ and $\mu$ such that chiral symmetry is essentially
restored. Our most realistic treatment of quarks uses a Nambu-Jona
Lasinio four-fermion interaction to model chiral symmetry breaking
effects, so these models are of Polyakov-Nambu-Jona Lasinio (PNJL)
type \cite{Fukushima:2003fw}.

\begin{table}
\begin{tabular}{|c|c|c|}
\hline 
Model of confinement & Confining potential $V_{d}$ & Parameter\tabularnewline
\hline 
\hline 
A & Eq. (\ref{eq:VA_d}) & $M_{A}=596$ MeV\tabularnewline
\hline 
B & Eq. (\ref{eq:VB_d}) & $R=1.0028$fm\tabularnewline
\hline 
\end{tabular}

\caption{\label{tab:confinement_parameters}Potential term and parameters for
modeling confinement effects. Parameters are determined from the deconfinement
temperature for pure $SU(3)$ gauge theory.}
\end{table}

\begin{table}
\begin{tabular}{|c|c|c|c|c|}
\hline 
Model of $N_{f}=2$ fermions & Quark potential $V_{f}$ & $m_{0}$ & $g_{S}$ & $\Lambda$\tabularnewline
\hline 
\hline 
Heavy Quarks & Eq. (\ref{Vf_T}) & $2000$ MeV & 0 & -\tabularnewline
\hline 
Massless Quarks & Eq. (\ref{eq:Vf_massless}) & 0 & 0 & -\tabularnewline
\hline 
PNJL & Eq. (\ref{eq:Vf_zero}) + Eq. (\ref{Vf_T}) & $5.5$ MeV & $5.496$ $\mbox{GeV}^{-2}$ & $631.4$ MeV\tabularnewline
\hline 
\end{tabular}

\caption{\label{tab:chiral_parameters}Potential term and parameters for quark
sector. All numerical values are for two-flavor QCD.}
\end{table}

In our PNJL models, we write the fermionic part of the partition function
as
\begin{equation}
Z_{f}=\int\mathcal{D}\bar{\psi}\mathcal{D}\psi e^{i\int d^{4}x\,\mathcal{L}_{f}}\label{eq:Z_psi}
\end{equation}
using $N_{f}=2$ Nambu-Jona-Lasinio-type Lagrangian with the constant
Polyakov loop \cite{Fukushima:2003fw} 
\begin{equation}
\mathcal{L}_{f}=\bar{\psi}(i\gamma\cdot D-m_{0})\psi+g_{S}\left\{ \left(\bar{\psi}\psi\right)^{2}+\left(\bar{\psi}i\gamma_{5}\lambda^{a}\psi\right)^{2}\right\} \label{eq:L_NJL}
\end{equation}
where $m_{0}$ is the current mass of the quarks, $g_{S}$ is the
four-fermion coupling, and $\lambda^{a}$'s are the generators of
the flavor symmetry group $SU(2)$. The covariant derivative $D_{\mu}$
couples the fermions to a background Polyakov loop via the component
of the gauge field in the temporal direction. Introducing auxiliary
fields, a scalar field $\sigma$ and triplet of pseudoscalar fields
$\pi^{a}$,
\begin{equation}
\mathcal{L}_{aux}=-g_{S}\left\{ \sigma^{2}+\left(\pi^{a}\right)^{2}\right\} +2g_{S}\bar{\psi}\left\{ \sigma+i\gamma_{5}\lambda^{a}\pi^{a}\right\} \psi,
\end{equation}
and integrating over the fermion fields, we can write the partition
function in terms of the boson fields (i.e. bosonization)
\begin{equation}
Z_{f}=\int\mathcal{D}\sigma\mathcal{D}\pi^{a}\exp\left[i\int d^{4}x\left\{ \mbox{tr}\log\left[i\gamma\cdot D-m_{0}+2g_{S}(\sigma+i\pi^{a}\lambda^{a})\right]-g_{S}\left(\sigma^{2}+(\pi^{a})^{2}\right)\right\} \right].
\end{equation}
 We use the background field method for the scalar field, $\sigma(x)=\sigma_{0}+s(x)$
and write the partition function as
\begin{equation}
Z_{f}=\exp\left[i\int d^{4}x\left\{ \mbox{tr}\log\left[i\gamma\cdot D-m\right]-g_{S}\sigma_{0}^{2}\right\} \right]\int\mathcal{D}s\mathcal{D}\pi^{a}\exp\left[i\int d^{4}x\,\mathcal{L}_{b}\right]\label{eq:Z_sigma}
\end{equation}
where $m=m_{0}-2g_{S}\sigma_{0}$ is the constituent quark mass m
, $\mbox{tr}$ denotes the trace over the color, flavor, and Dirac
space, and the bosonized Lagrangian is 
\begin{equation}
\mathcal{L}_{b}=\mbox{tr}\log\left[1+\frac{1}{i\gamma\cdot D-m}2g_{S}(s+i\pi^{a}\lambda^{a})\right]-g_{S}\left(s^{2}+(\pi^{a})^{2}\right).\label{L_b}
\end{equation}

We perform a Wick rotation and consider the theory in Euclidean space
from now on. The first term in the partition function (\ref{eq:Z_sigma})
gives the effective potential,
\begin{equation}
V_{f}=V_{f}^{T}(P,m)+V_{f}^{0}(m,m_{0}),
\end{equation}
which consists of the finite-temperature part $V_{f}^{T}$, which
is given by Eq. (\ref{eq:Vf_def_1loop}) and the vaccum part $V_{f}^{0}$,
\begin{equation}
V_{f}^{0}(m,m_{0})=\frac{\left(m-m_{0}\right)^{2}}{4g_{S}}-2N_{f}\mbox{Tr}_{F}\int\frac{d^{3}k}{\left(2\pi\right)^{3}}\omega_{k}.
\end{equation}
We note that the finite-temperature contribution $V_{f}^{T}$ is finite
for any values of $P$, $m$, $\mu$, and $T$, while the zero-point
energy, the integral in $V_{f}^{0}$, is divergent and needs a regularization.
We use a noncovariant three-dimensional cutoff, $\Lambda$ \cite{Klevansky:1992qe}
and write it as \cite{Nishimura:2009me} 
\begin{eqnarray}
V_{f}^{0}(m,m_{0}) & = & \frac{\left(m-m_{0}\right)^{2}}{4g_{S}}-\frac{N_{c}N_{f}\Lambda^{4}}{8\pi^{2}}\left\{ \sqrt{1+\left(m/\Lambda\right)^{2}}\left[2+\left(m/\Lambda\right)^{2}\right]\right.\nonumber \\
 &  & \left.+(m/\Lambda)^{4}\log\frac{m/\Lambda}{1+\sqrt{1+\left(m/\Lambda\right)^{2}}}\right\} .\label{eq:Vf_zero}
\end{eqnarray}
For $V_{f}^{T}$, it is often convenient to combine the arguments
of the logarithms into a single product that is manifestly real. Using
Eq. (\ref{eq:Vf_def_1loop}), we can write the finite-temperature
effective potential in terms of Polyakov loop eigenvalues as
\begin{eqnarray}
V_{f}^{T} & = & -2TN_{f}\sum_{j=1}^{N_{c}}\int\frac{d^{3}k}{\left(2\pi\right)^{3}}\left[\log\left(1+e^{-\left(\omega_{k}-\mu\right)/T+i\theta_{j}}\right)+\log\left(1+e^{-\left(\omega_{k}+\mu\right)/T-i\theta_{j}}\right)\right]\nonumber \\
 & = & -2TN_{f}\int\frac{d^{3}k}{\left(2\pi\right)^{3}}\left\{ \log\left[1+2\cos\theta\, e^{-\left(\omega_{k}-\mu\right)/T+\psi}+e^{-2\left(\omega_{k}-\mu\right)/T+2\psi}\right]\right.\label{eq:}\\
 &  & \left.+\log\left[1+e^{-\left(\omega_{k}-\mu\right)/T-2\psi}\right]+\left(z\rightarrow-z\right)\right\} \label{Vf_T}
\end{eqnarray}
where the last part denotes the antiparticle contribution which has
the opposite sign for the chemical potential and the Polyakov loop
eigenvalues, $z=\mu-igA_{\mu}$. From this expression, we can see
explicitly that the one-loop fermionic effective potential at the
complex saddle point is real, independent of any approximation. We
use Eqs. (\ref{eq:Vf_zero}) and (\ref{Vf_T}) for the effective potential
of the fermionic part of PNJL model with the $T=0$ parameters taken
from \cite{Hatsuda:1994pi}. 

In principle, the coupling of $P$ and $\bar{\psi}\psi$ which is
a prominent feature of PNJL models can lead to an extended mass matrix
that incorporates mixing of $\bar{\psi}\psi$ with excitations of
the Polyakov loop. The kinetic term of the scalar field $s$ in the
bosonized Lagrangian is needed for a full treatment. Using the log
expansion and the derivative expansion for Eq. (\ref{L_b}) \cite{Eguchi:1976iz,Klevansky:1992qe},
we can obatin the kinetic term for the scalar field in the form
\[
\mathcal{L}_{b}\supset4N_{f}g_{S}^{2}\mbox{Tr}_{F}I_{s}^{\mu\nu}\partial_{\mu}s\partial_{\nu}s
\]
where $I_{s}^{\mu\nu}$ is, for example, given as the momentum integral
in Eq. (7.54) of \cite{Klevansky:1992qe} but the four-momentum $k_{\mu}$
is replaced by $k_{\mu}+gA_{4}\delta_{\mu4}$ for the PNJL model.
Using the identity
\[
\int\frac{d^{3}k}{(2\pi)^{3}}k_{i}k_{j}=\int\frac{d^{3}k}{(2\pi)^{3}}\frac{k^{2}\delta_{ij}}{3}
\]
and rescaling $s$ for the physical constituent mass, we can write
the spatial part of the kinetic term as
\[
\mathcal{L}_{b}\supset\frac{1}{2}I_{s}\left[\partial_{i}\left(-2g_{S}s\right)\right]^{2}
\]
with
\begin{eqnarray}
I_{s} & = & N_{f}T\sum_{n=-\infty}^{\infty}\int\frac{d^{3}k}{(2\pi)^{3}}\mbox{Tr}_{F}\left[\left\{ \frac{2}{\left[\left(\omega_{n}+iz\right)^{2}+\omega_{k}^{2}\right]^{2}}-\frac{\frac{4}{3}k^{2}+4m^{2}}{\left[\left(\omega_{n}+iz\right)^{2}+\omega_{k}^{2}\right]^{3}}\right.\right.\label{eq:Is}\\
 &  & \left.\left.+\frac{\frac{16}{3}k^{2}m^{2}}{\left[\left(\omega_{n}+iz\right)^{2}+\omega_{k}^{2}\right]^{4}}\right\} +(z\rightarrow-z)\right]\nonumber 
\end{eqnarray}
where $z=\mu-igA_{4}$ and the summation is over the Matsubara frequencies,
$\omega_{n}=(2n+1)\pi$. A similar expression for $I_{s}$ is obtained
in \cite{Imai:2012hr} for the PNJL model. We first use the prescription
(\ref{eq:EVs_PolyakovLoop_SU3}) for the Polyakov loop and sum over
the Matsubara frequencies and integrate over the three-momentum in
Eq. (\ref{eq:Is}). However, the integral is divergent, and we use
the same non-covariant three-dimensional cutoff $\Lambda$ used for
the zero-point energy (\ref{eq:Vf_zero}). With the $s$ kinetic term
given in terms of $I_{s}$, we can in principle compute the eigenvalues
of an extended mass matrix. It turns out, however, that the off-diagonal
coupling of the chiral component of the mass matrix is numerically
negligible compared to the Polyakov-loop parts of the mass matrix,
and thus we ignore the chiral component in the remainder of this paper.

\section{Heavy quarks}

We consider the case of heavy quarks propagating in constant Polyakov
loop background. For such quark, the chiral symmetry effects are negligible
and a first-order deconfinement transition line is the only true critical
behavior found in the phase diagram. Our study of heavy quarks is
perhaps most relevant for lattice studies of static quarks at non-zero
$\mu$; this approximation is particularly tractable \cite{Blum:1995cb}.

\begin{figure}
\includegraphics[width=5in]{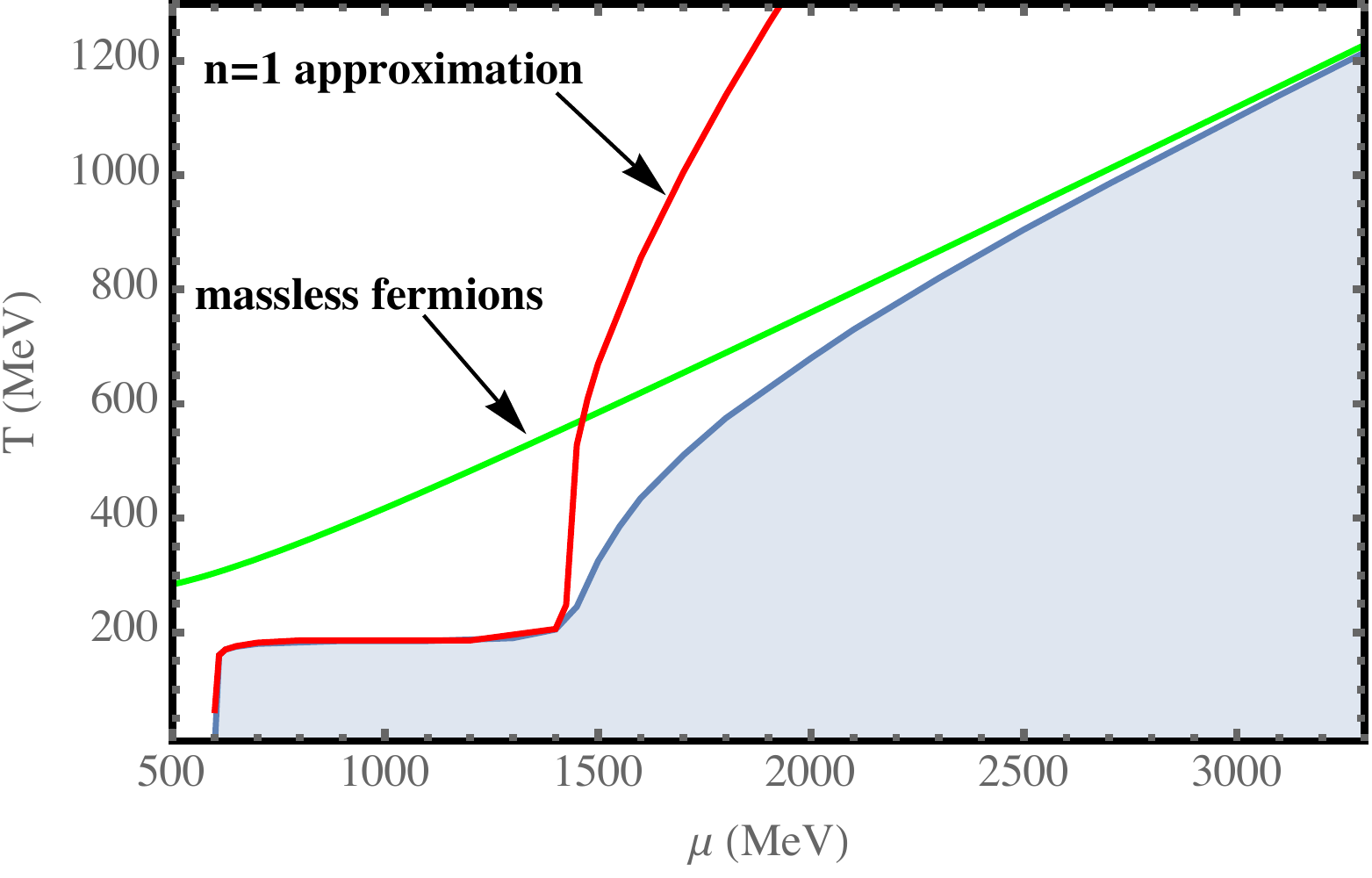}

\caption{\label{fig:hq_mm_modelA}The shaded$ $ region indicates where $\kappa_{I}\ne0$
for heavy quarks ($m=2000\,\mbox{MeV)}$ using Model A for confinement
effects. The boundary of this region is also shown using an approximation
appropriate for very heavy quarks ($\beta m\gg1$) as well as for
massless quarks, appropriate when $\beta m\ll1$.}
\end{figure}

The center symmetry of pure gauge theory is exact for infinitely heavy
quarks. However, quarks with finite mass break the center symmetry
explicitly and weaken the first order transition of pure gauge theory.
At sufficiently low quark mass the first order transition for deconfinement
vanishes at a critical end point. The location of this critical end
point is model dependent and has been proposed as a useful way to
differentiate between different models of confinement \cite{Kashiwa:2012wa}.
In both Model A and Model B, the first order deconfinement transition
vanishes for quark mass of around 1.5 GeV or less. Therefore we set
the quark mass to be 2 GeV so that the deconfinement transition still
persists. The end point of the deconfinement transition line lies
at smaller values of $\mu$, and appears to play no direct role in
the behavior of $\psi$ and $\kappa_{I}$. The quark mass is large
compared to the confinement scale, so asymptotic freedom applies in
the region $\mu\simeq m$. In this region, perturbation theory is
a reliable guide when $T\gg T_{d}$, the pure gauge deconfinement
temperature. However, below $T_{d}$, non-perturbative confinement
effects cannot be neglected, hence the importance of the potential
term $V_{d}$ beyond what is usually considered the confining region
at low $T$ and $\mu$. A useful expansion for $\beta\left(m-\mu\right)\gg1$
for can be generated by expanding the logarithm in Eq. (\ref{eq:Vf_def_1loop})
and integrating term by term \cite{Meisinger:2001fi}. Such an expansion
gives
\begin{equation}
V_{f}\left(P\right)=\sum_{n=1}^{\text{\ensuremath{\infty}}}\frac{(-1)^{n}m^{2}T^{2}}{n^{2}\text{\textgreek{p}}^{2}}K_{2}(n\beta m)(e^{\frac{n\mu}{T}}{\rm Tr}_{F}P^{n}+e^{-\frac{n\mu}{T}}{\rm Tr}_{F}P^{\dagger n}).\label{eq:Vf_hq}
\end{equation}
At low temperature and density the effects of heavy quarks can be
obtained approximately from the $n=1$ term of Eq. (\ref{eq:Vf_hq}).
However, this expansion fails in the high density region ($\mu>1.5$
GeV) in case of Model A as can be seen in Fig. \ref{fig:hq_mm_modelA}.
In our analysis we have therefore numerically integrated the full
one loop expression for heavy quark potential. 

\begin{figure}
\includegraphics[width=5in]{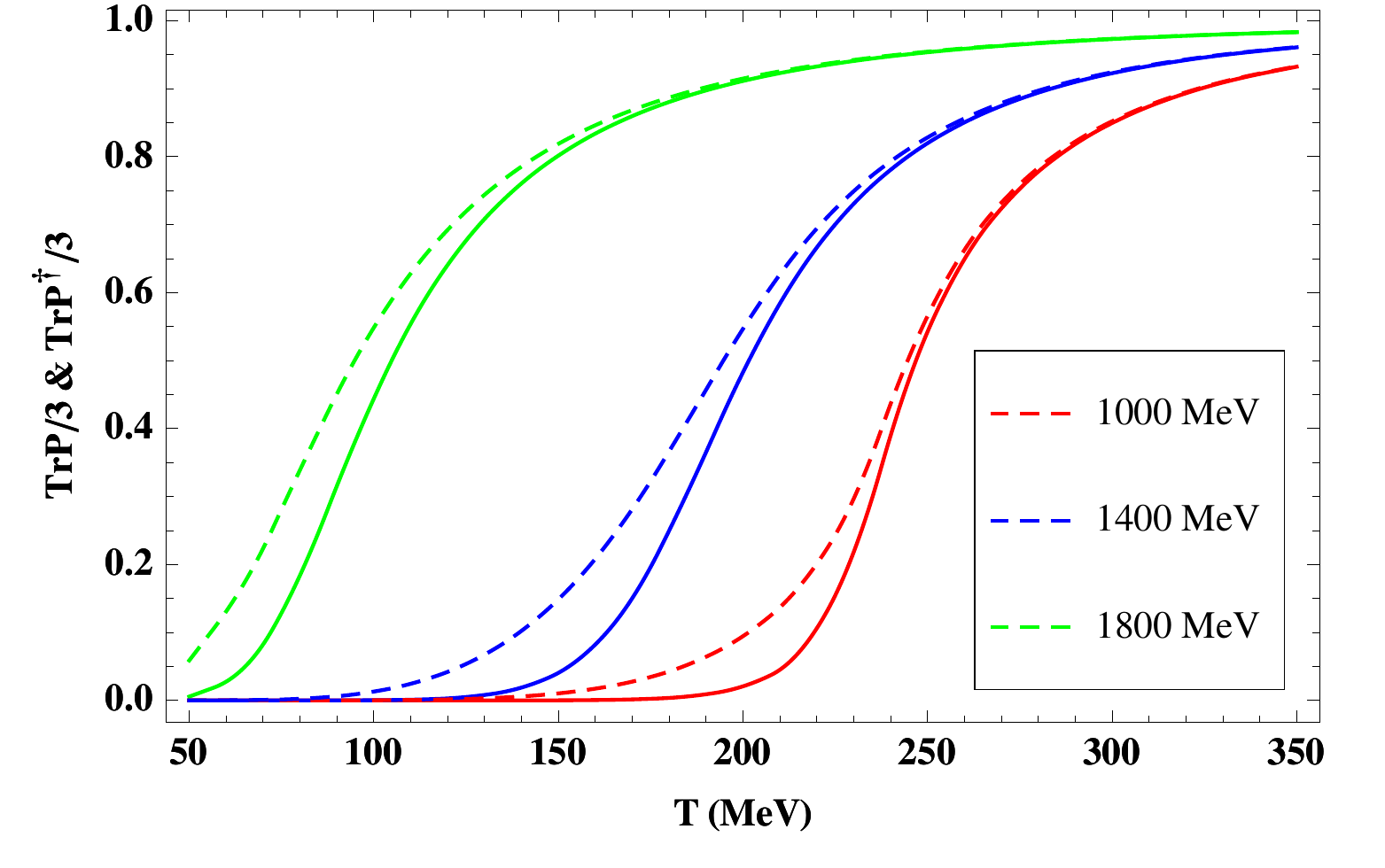}

\caption{\label{fig:hq_modelA_TrP}$\left\langle \mbox{Tr}_{F}P\right\rangle $
and $\left\langle \mbox{Tr}_{F}P^{\dagger}\right\rangle $ as a function
of $T$ for $\mu=1000,\,1400$ and $1800$ MeV for heavy quarks using
Model A for confinement effects. The Polyakov loops are normalized
to one as the temperature becomes large.}
\end{figure}

\begin{figure}
\includegraphics[width=5in]{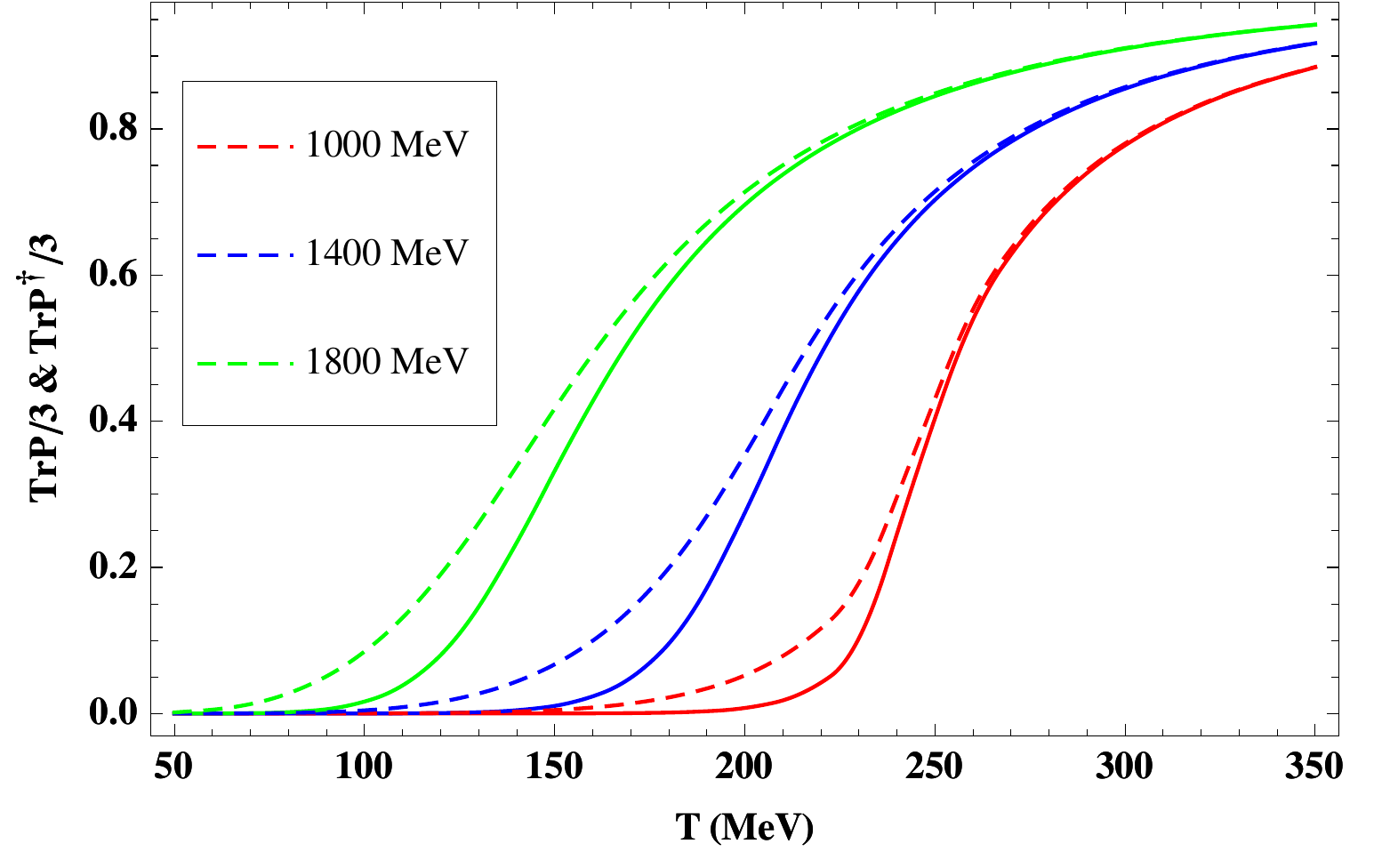}

\caption{\label{fig:hq_modelB_TrP}$\left\langle \mbox{Tr}_{F}P\right\rangle $
and $\left\langle \mbox{Tr}_{F}P^{\dagger}\right\rangle $ as a function
of $T$ for $\mu=1000,\,1400$ and $1800$ MeV for heavy quarks using
Model B for confinement effects. The Polyakov loops are normalized
to one as the temperature becomes large.}
\end{figure}

In Fig. \ref{fig:hq_modelA_TrP} we show $\mbox{Tr}_{F}P$ and $\mbox{Tr}_{F}P^{\dagger}$
as a function of $T$ for various values of $\mu$ when the heavy
quark has a mass of $2\,\mbox{GeV}$. In aggreement with our general
argument above, the crossover moves toward lower values of $T$ as
$\mu$ increases. The separation between $\mbox{Tr}_{F}P$ and $\mbox{Tr}_{F}P^{\dagger}$
is largest in the crossover region, and is negligible at higher temperatures.
As shown in the figure, the separation is largest for some intermediate
value of $\mu$ less than the heavy quark mass. The behavior of $\mbox{Tr}_{F}P$
and $\mbox{Tr}_{F}P^{\dagger}$ for Model B is similar to Model A,
as may be seen from Fig. \ref{fig:hq_modelB_TrP}. The crossover happenes
at higher temperature for Model B, showing that the confining effect
for Model B is smaller than Model A. Because $\psi$ is non-zero in
both models, there is a difference between $\mbox{Tr}_{F}P$ and $\mbox{Tr}_{F}P^{\dagger}$. 

\begin{figure}
\includegraphics[width=5in]{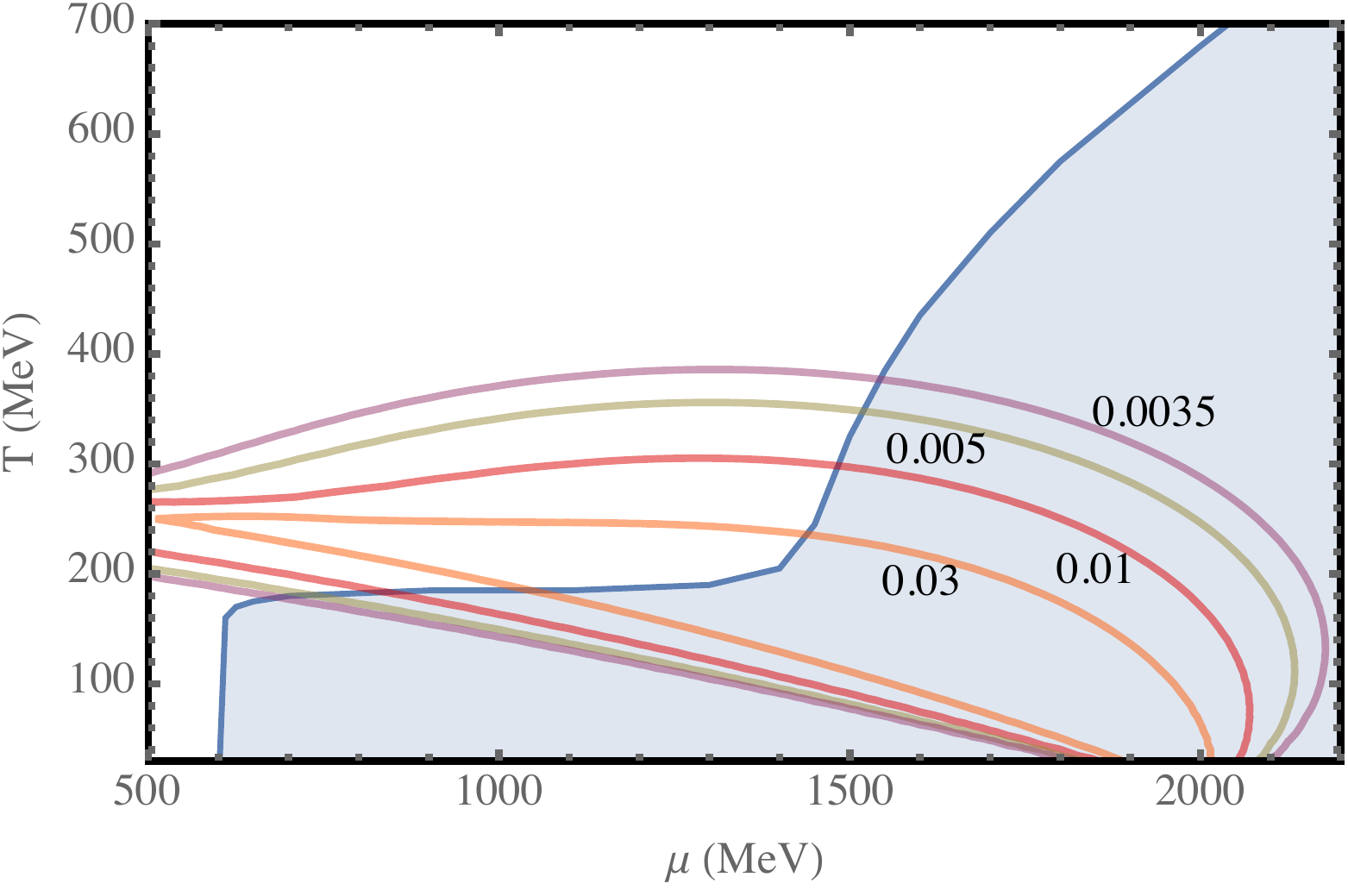}

\caption{\label{fig:hq_modelA_contours_psi}Contour plot of $\psi$ in the
$\mu-T$ plane for heavy quarks ($m=2000\,\mbox{MeV)}$ using Model
A for confinement effects. The region where $\kappa_{I}\ne0$ is shaded. }
\end{figure}

In Fig. \ref{fig:hq_modelA_contours_psi}, we show for Model A a contour
plot for $\psi$ along with a shaded region showing where $\kappa_{I}\ne0$.
The boundary of the shaded region is thus the disorder line. From
this graph, we see that values of $\psi$ are very small, but peak
in a region centered roughly around $\mu=1500$ MeV and $T=150$ MeV.
There is no obvious relation between the region where $\psi$ is largest
and the region where $\kappa_{I}\ne0$ . However, the peak in $\psi$
is located near the point where the disorder line abruptly changes.

\begin{figure}
\includegraphics[width=5in]{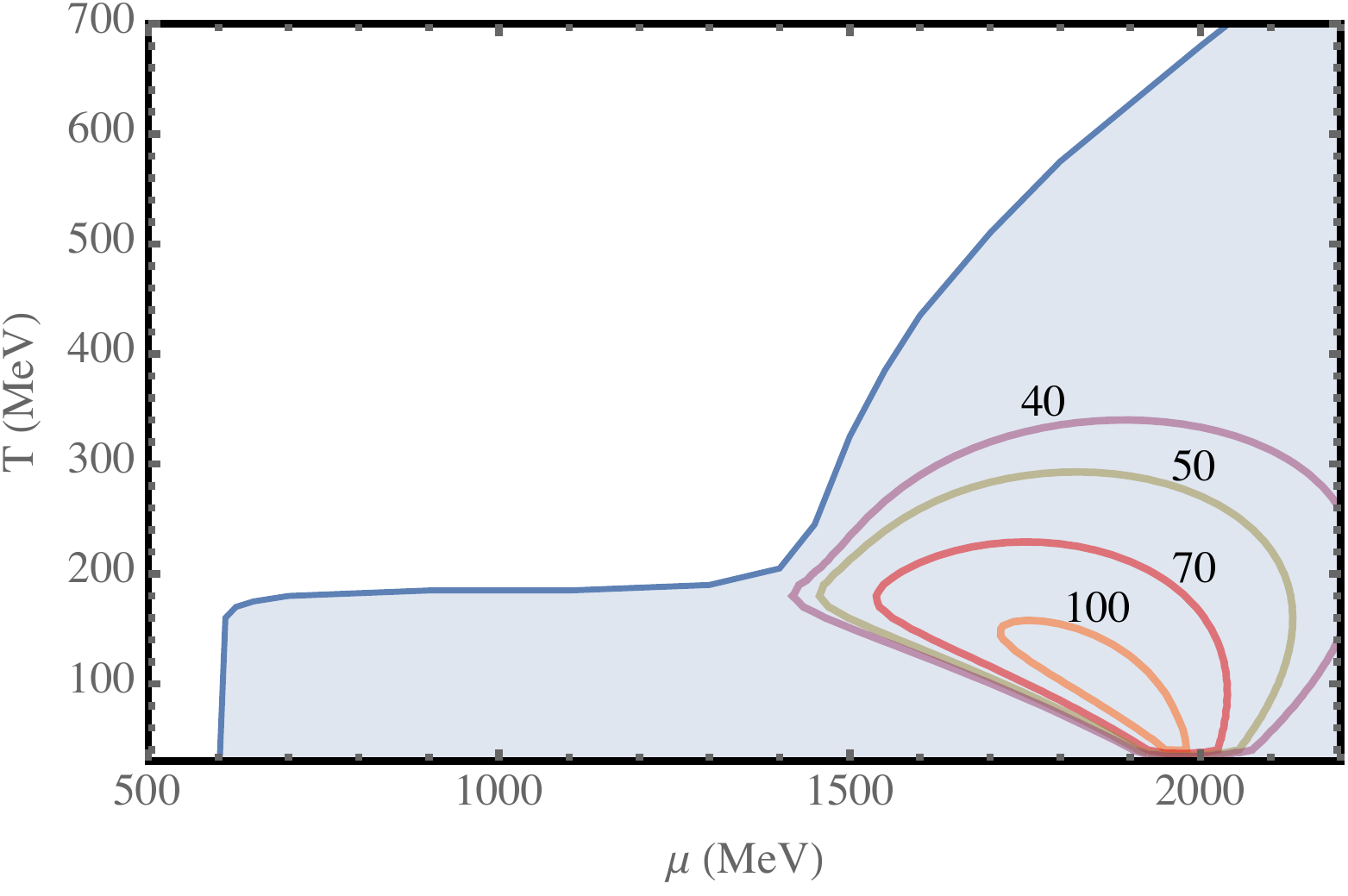}

\caption{\label{fig:hq_modelA_contours_mm}Contour plot of $\kappa_{I}$ in
the $\mu-T$ plane for heavy quarks ($m=2000\,\mbox{MeV)}$ using
Model A for confinement effects. Contours are given in MeV with $\alpha_{S}$
set to one. The region where $\kappa_{I}\ne0$ is shaded. }
\end{figure}

Figure \ref{fig:hq_modelA_contours_mm} again shows the region where
$\kappa_{I}\ne0$ and the associated disorder line, but now with contour
lines for $\kappa_{I}$ added. As with all the contour plots of this
type, we have set the running coupling $\alpha_{s}\left(T,\mu\right)=1$.
In other words, conversion to the actual one-loop values requires
multiplying these values by appropriate values for $\sqrt{\alpha_{s}\left(T,\mu\right)}$.
The region where the mass eigenvalues $\kappa_{1}$ and $\kappa_{2}$
form a complex conjugate pair has a complicated shape. The mass matrix
eigenvalues are real for $\mu$ below about $600\,\mbox{MeV}$. There
is a roughly rectangular region for $600\,\mbox{MeV}\lesssim\mu\lesssim1450\,\mbox{MeV}$
. This is followed by a region where the boundary rises roughly linearly
with $\mu$ , similar to the behavior of Model A with massless quarks.

Figure \ref{fig:hq_mm_modelA} shows the physics associated with this
behavior. The boundary using the complete one-loop expression is compared
with both the massless boundary and the boundary obtained using the
$n=1$ approximation from Eqn. \ref{eq:Vf_hq} to the full one-loop
expression. As may be seen, the $n=1$ term accounts very well for
the low-temperature behavior of the boundary, while the massless quark
result is accurate for $\mu$ above the heavy quark mass. It is clear
that the abrupt change of the shaded region represents a rapid crossover
from the behavior of a heavy quark to the behavior of a massless quark,
occuring over a range of roughly $3M/4<\mu<5M/4$ , with most of the
change occuring before $\mu$ reaches $M$ .

\begin{figure}
\includegraphics[width=5in]{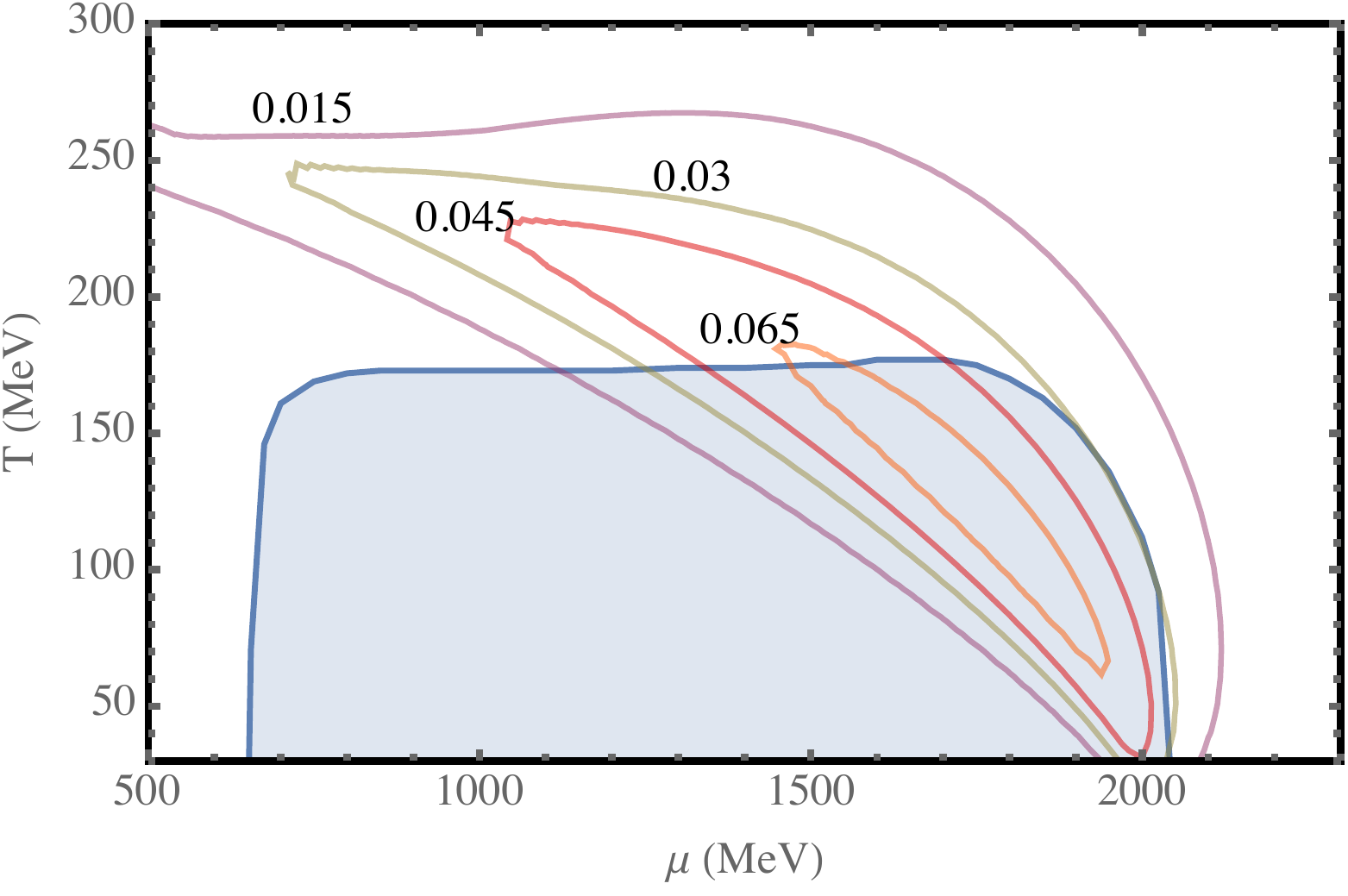}

\caption{\label{fig:hq_modelB_contours_psi}Contour plot of $\psi$ in the
$\mu-T$ plane for heavy quarks ($m=2000\,\mbox{MeV)}$ using Model
B for confinement effects. The region where $\kappa_{I}\ne0$ is shaded. }
\end{figure}

\begin{figure}
\includegraphics[width=5in]{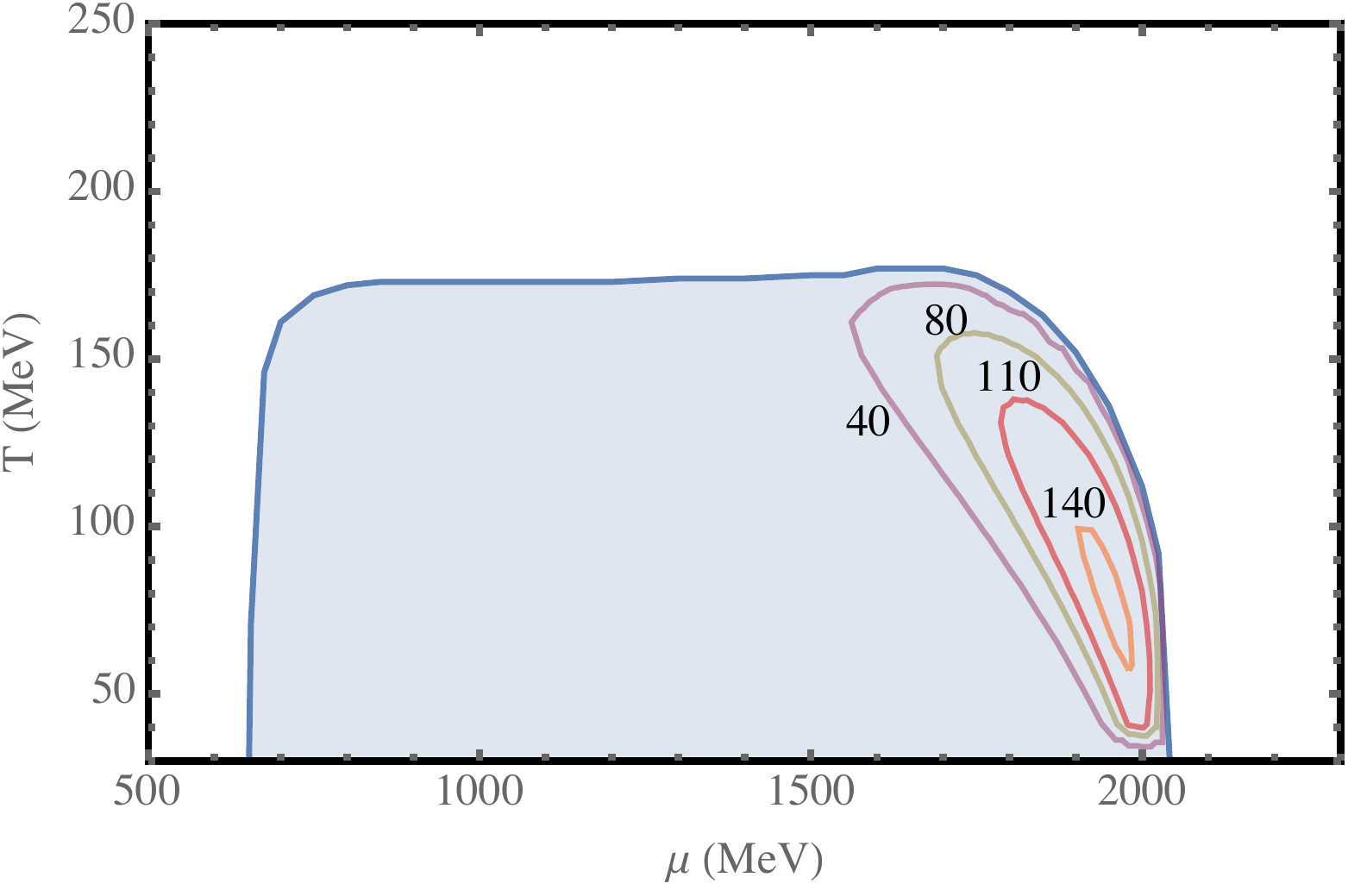}

\caption{\label{fig:hq_modelB_contours_mm} Contour plot of $\kappa_{I}$ in
the $\mu-T$ plane for heavy quarks ($m=2000\,\mbox{MeV\ensuremath{}}$)
using Model B for confinement effects. Contours are given in MeV with
$\alpha_{S}$ set to one. The region where $\kappa_{I}\ne0$ is shaded. }
\end{figure}

The behavior of $\psi$ for Model B, as shown in Figure\ \ref{fig:hq_modelB_contours_psi},
is similar to the behavior of $\psi$ for Model A, but the values
of $\psi$ are somewhat larger. The region where the eigenvalues of
the mass matrix are complex is shown in Fig. \ref{fig:hq_modelB_contours_mm}.
The shape and size of the region is very similar to the rectangular
region found for Model A in Fig.\ \ref{fig:hq_modelA_contours_mm}.
However, in the high-temperature region, where $\mu$ is greater than
the quark mass, the region of complex mass eigenvalues is completely
missing for Model B. This is consistent with the behavior of Model
B for massless quarks, where no complex eigenvalues of the mass matrix
were found.

\begin{figure}
\includegraphics[width=5in]{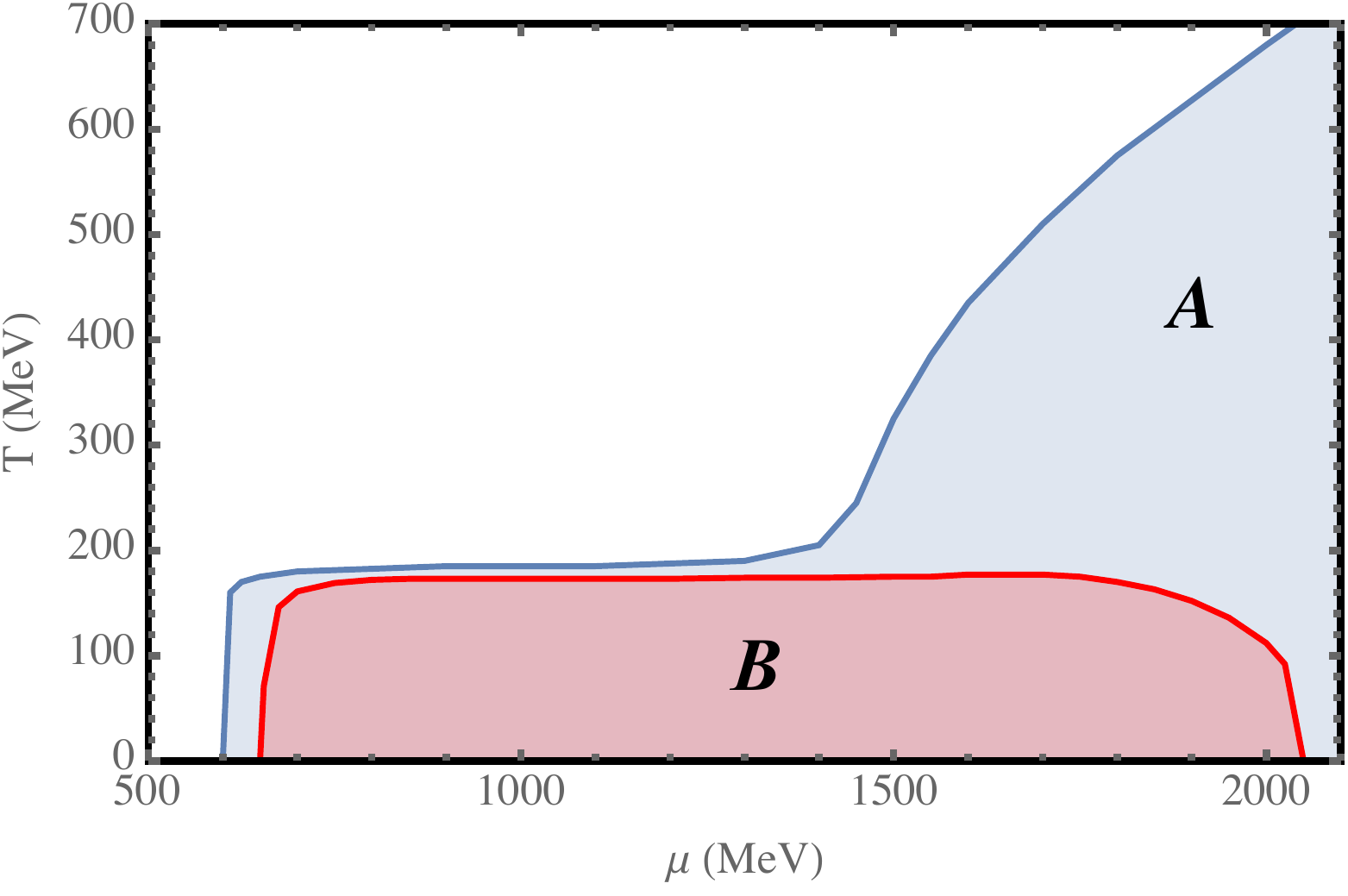}

\caption{\label{fig:Comparison-AB-hq}A comparison of the regions where $\kappa_{I}\ne0$
for heavy quarks with Model A and Model B along with the corresponding
disorder lines.}

\end{figure}

Figure \ref{fig:Comparison-AB-hq} shows a comparison of the regions
where $\kappa_{I}$ is non-zero for both Model A and Model B. Their
shape is very similar for smaller values of $\mu$, suggesting that
some universal behavior occurs in this region. However, the behavior
is very different in the region where both $T$ and $\mu$ are becoming
large. Model A shows a continuation of the disorder line that follows
the behavior for massless quarks, while for Model B the disorder line
covers a finite region in $\mu-T$ space.

\section{Massless quarks without chiral effects}

In this section we extend the results of our previous work on massless
quarks using Model A \cite{Nishimura:2014rxa}, including more detail
and providing a comparison with Model B. This simple model where the
quark mass $m$ is set to zero neglects chiral symmetry breaking,
relevant at low $T$ and low $\mu$. It should not be expected to
reproduce exactly the features seen in lattice simulations. Nevertheless,
comparison with PNJL model results, \emph{e.g.}, \cite{Schaefer:2007pw},
show that the model is quantitatively similar to the behavior of models
with many more free parameters that include chiral symmetry effects.
For Model A, $\mathrm{Tr}_{F}P$ shows a slightly larger initial rise
in $\mathrm{Tr}_{F}P$ with temperature than does the model studied
in \cite{Schaefer:2007pw}. This is consistent with the role that
chiral symmetry breaking plays in diminishing the explicit breaking
of $Z(3)$ symmetry by quarks.

\begin{figure}
\includegraphics[width=5in]{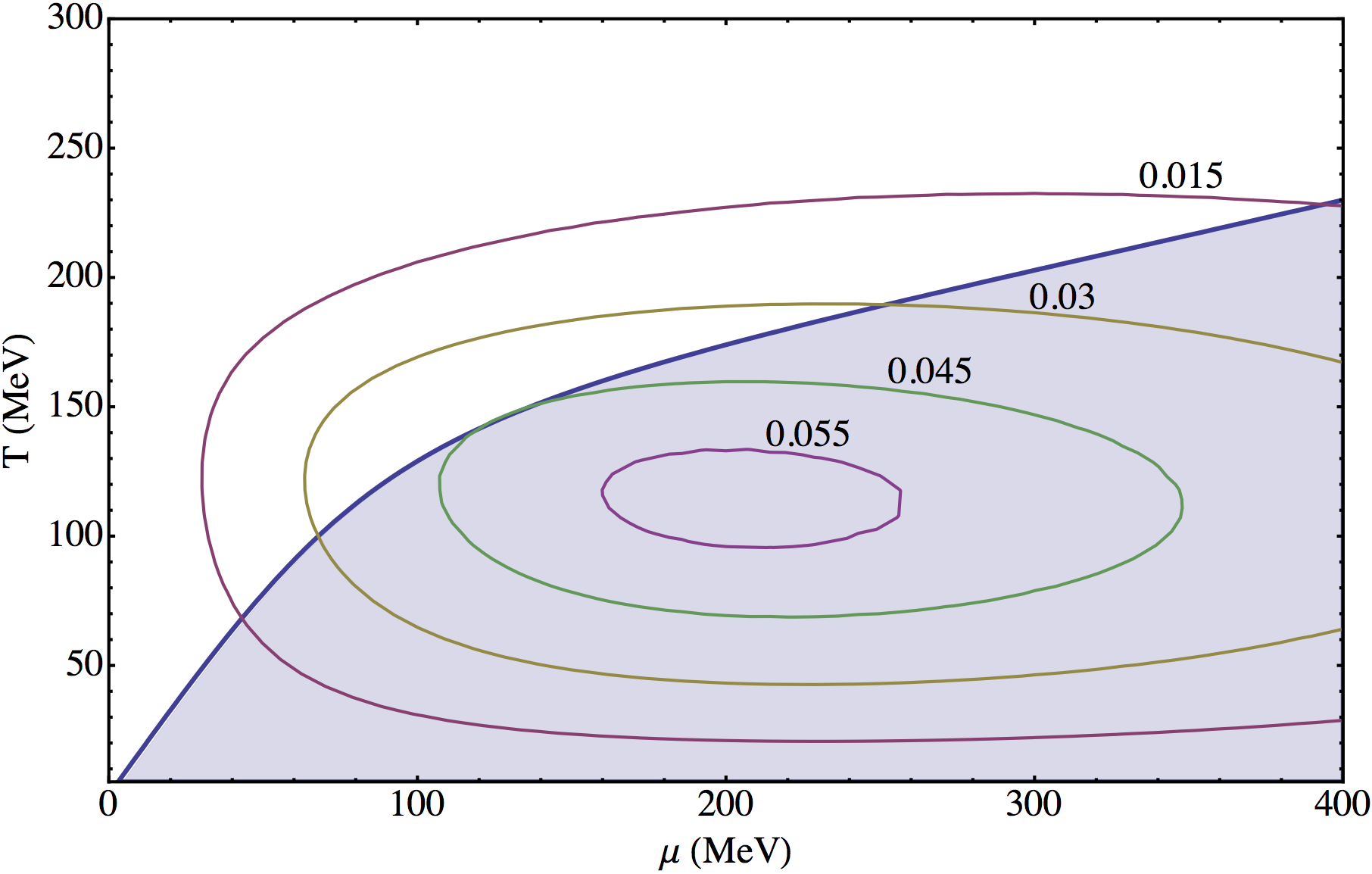}

\caption{\label{fig:Light_A_Psi_DisorderLine}Contour plot of $\psi$ in the
$\mu-T$ plane for Model A with massless quarks, showing where ${\rm Tr}_{F}P$
is most different from ${\rm Tr}_{F}P^{\dagger}$. The region where
$\kappa_{I}\ne0$ is shaded.}
\end{figure}

\begin{figure}
\includegraphics[width=5in]{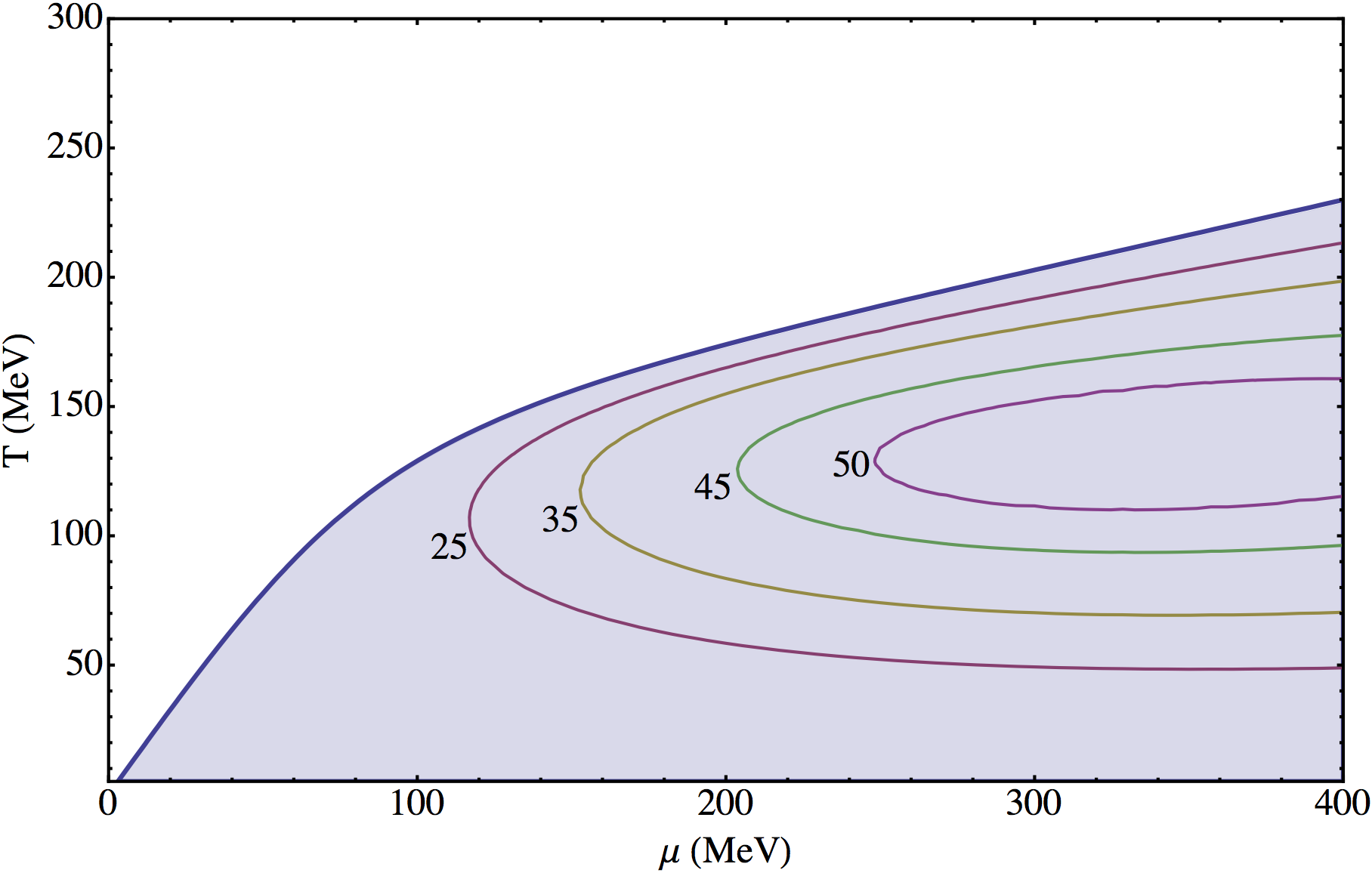}

\caption{\label{fig:Light_A_ImaginaryMass_DisorderLine}Contour plot of $\kappa_{I}$
in the $\mu-T$ plane for Model A with massless quarks. Contours are
given in MeV, with $\alpha_{s}$ set to one. The region where $\kappa_{I}\ne0$
is shaded.}
\end{figure}

\begin{figure}
\includegraphics[width=5in]{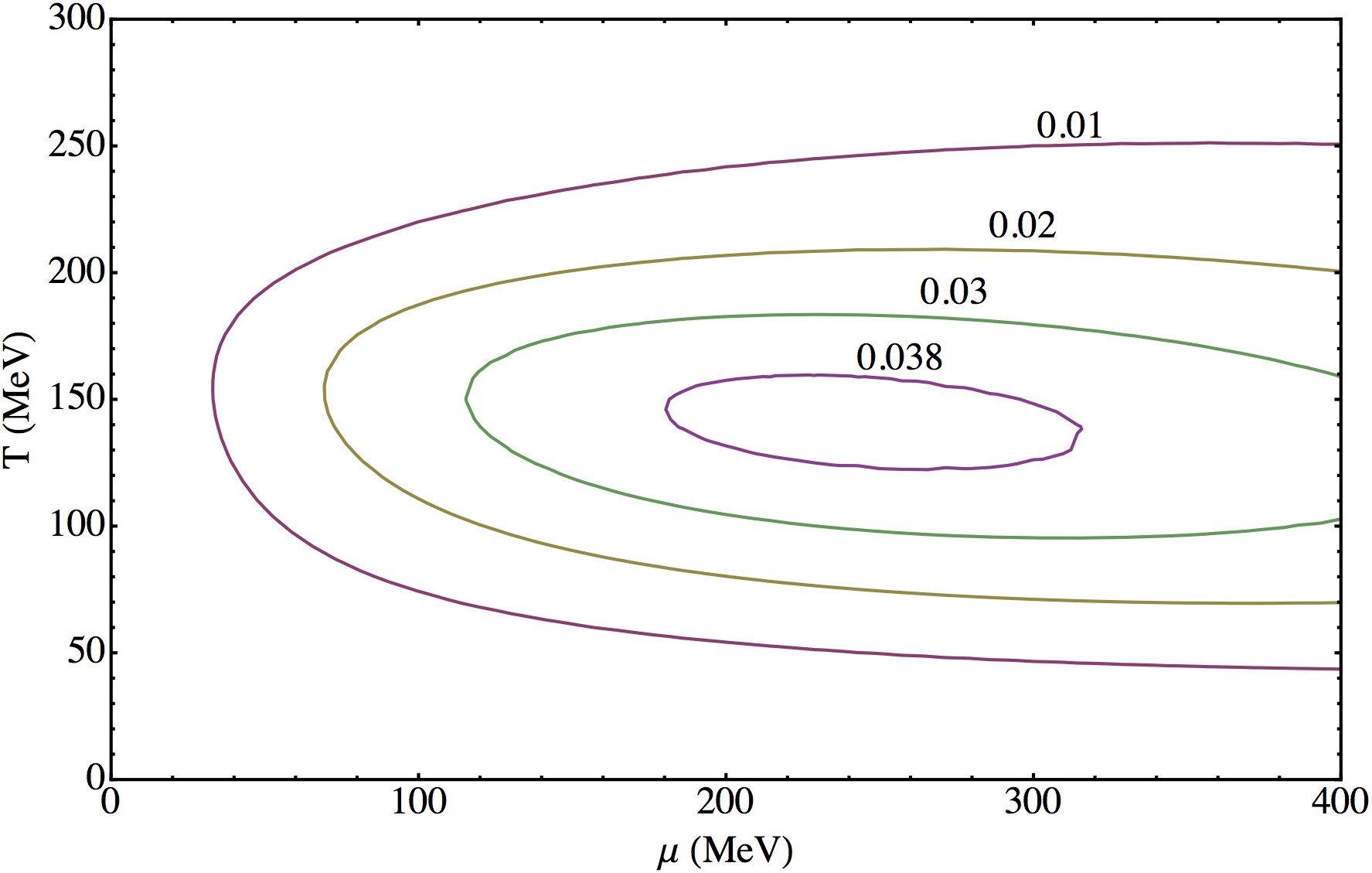}

\caption{\label{fig:Light_B_Psi}Contour plot of $\psi$ in the $\mu-T$ plane
for Model B with massless quarks, showing where ${\rm Tr}_{F}P$ is
most different from ${\rm Tr}_{F}P^{\dagger}$.}
\end{figure}

Figure \ref{fig:Light_A_Psi_DisorderLine} shows the region for Model
A where $\kappa_{I}$ is non-zero superimposed on a contour plot of
$\psi$, while \ref{fig:Light_A_ImaginaryMass_DisorderLine} shows
contour lines for $\kappa_{I}$. Comparison of the two figures shows
that the peak in $\psi$ occurs at a lower value of $\mu$ than the
peak in $\kappa_{I}$, with the peak in $\psi$ occuring near ($\mu=200$
MeV, $T=110$ MeV). The behavior of the disorder line for large $T$
and $\mu$ is known analytically \cite{Nishimura:2014rxa}:
\begin{equation}
T=\frac{2\mu}{\sqrt{3}\pi}.
\end{equation}
This behavior is generic to Model A when $T,\mu\gg m$, as we have
seen for heavy quarks in the previous section. 

The most interesting feature of Model B with massless quarks is that
there is no region where $\kappa_{I}$ is non-zero. Nevertheless,
as shown in Fig. \ref{fig:Light_B_Psi}, $\psi$ is non-zero, with
a peak value near ($\mu=250$ MeV, $T=140$ MeV). This is the only
case we have considered where there is no disorder line.

\section{PNJL models}

\begin{figure}
\includegraphics[width=5in]{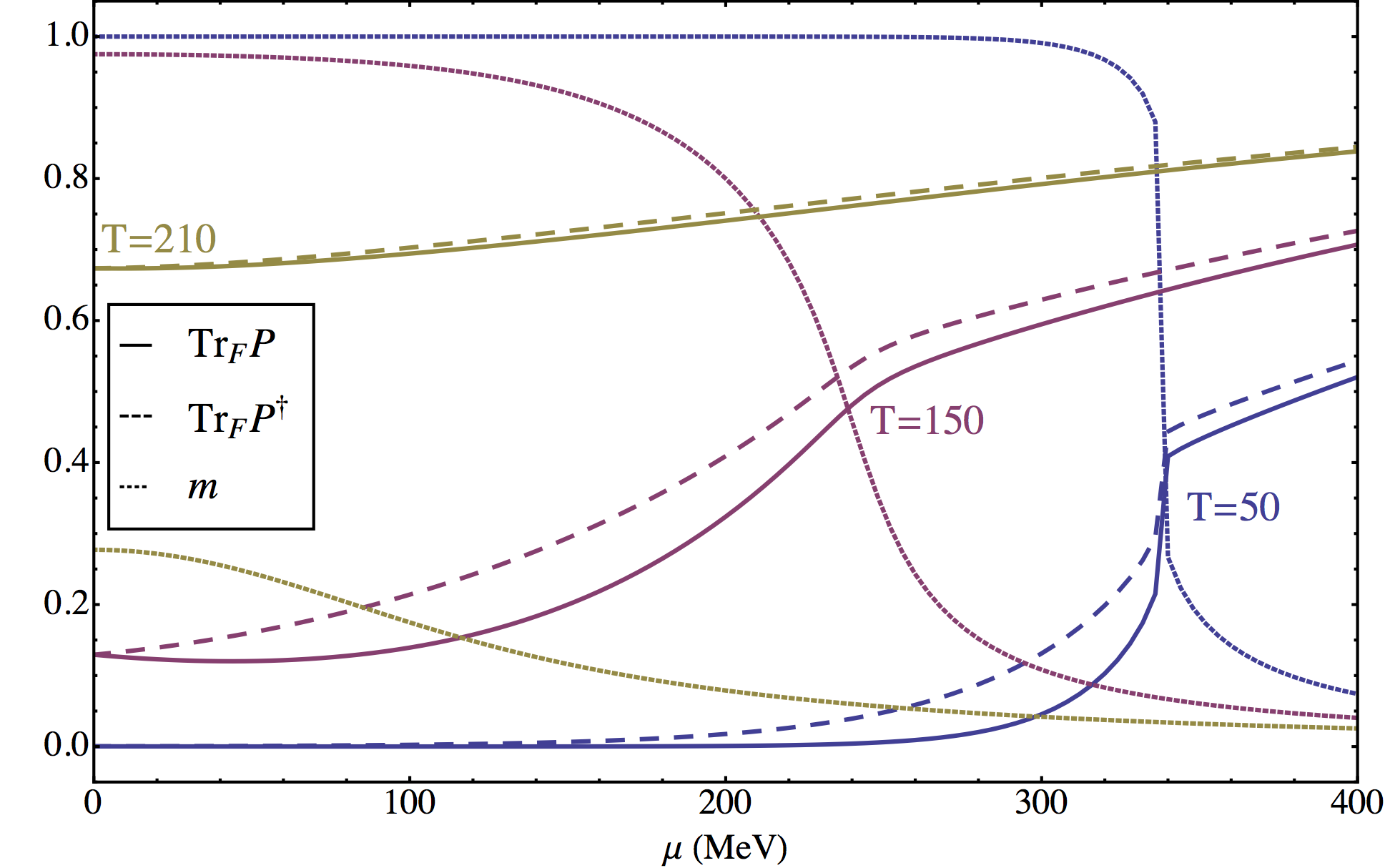}

\caption{\label{fig:OrderParameters_constantT_A}The constituent mass $m$,
$\left\langle {\rm Tr}_{F}P\right\rangle $ and $\left\langle {\rm Tr}_{F}P^{\dagger}\right\rangle $
as a function of $\mu$ for $T=50,\,150,\,\mbox{and \ensuremath{210}\,\mbox{MeV}}$
for a PNJL model using Model A for confinement effects. The constituent
mass $m$ is normalized to one at $T=0$, and the Polyakov loops are
normalized to one as the temperature becomes large.}
\end{figure}

\begin{figure}
\includegraphics[width=5in]{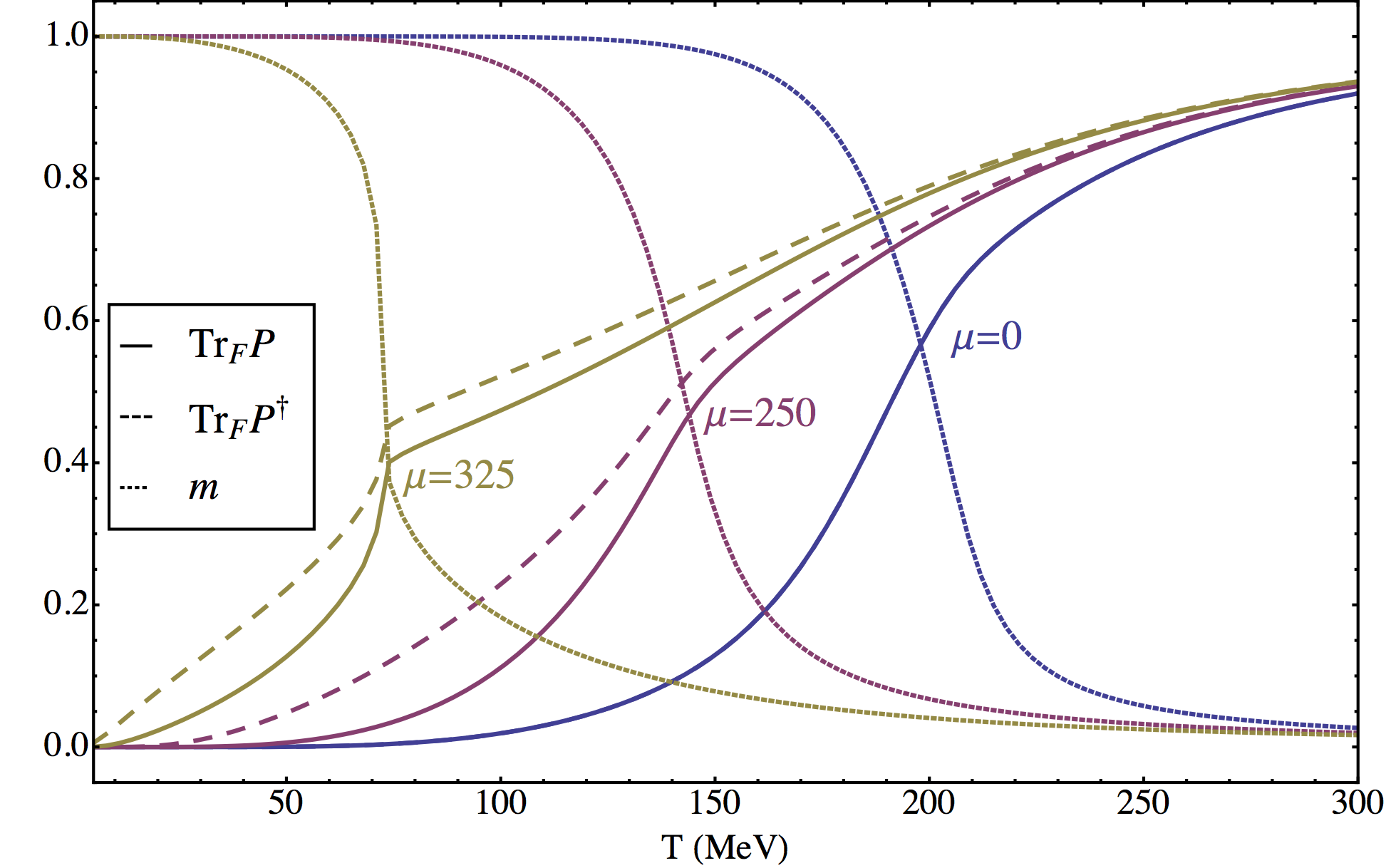}

\caption{\label{fig:OrderParameters_constantMu_A}The constituent mass $m$,
$\left\langle {\rm Tr}_{F}P\right\rangle $ and $\left\langle {\rm Tr}_{F}P^{\dagger}\right\rangle $
as a function of $T$ for $\mu=0,\,250$ and $325$ MeV for a PNJL
model using Model A for confinement effects. The constituent mass
$m$ is normalized to one at $T=0$, and the Polyakov loops are normalized
to one as the temperature becomes large.}
\end{figure}

In this section we consider our most realistic models of QCD at finite
temperature and density, PNJL models evaluated at complex saddle points.
These models have a much richer structure because the effects of chiral
symmetry breaking are included. Because the effective quark mass varies
with $T$ and $\mu$, the behavior of the PNJL models in some sense
lies between that of the heavy quarks and $m=0$ quarks considered
in the previous two sections, with a constituent quark mass that varies
with $T$ and $\mu$. Figures \ref{fig:OrderParameters_constantT_A}
and \ref{fig:OrderParameters_constantMu_A} show the values of $\left\langle {\rm Tr}_{F}P\right\rangle $,
$\left\langle {\rm Tr}_{F}P^{\dagger}\right\rangle $ and $m$ for
a PNJL model using $V_{d}^{A}$ to implement confinement. In all figures
of this type, the constituent quark mass $m$ is normalized to its
value at ($\mu=0$, $T=0$ ), while $\left\langle {\rm Tr}_{F}P\right\rangle $
and $\left\langle {\rm Tr}_{F}P^{\dagger}\right\rangle $ are normalized
so that they go to one as $T$ goes to infinity. As is typical of
PNJL models with appropriately chosen parameters, only crossover behavior
is seen at $\mu=0$. There is a critical line starting at $\mu\approx350$
MeV when $T=0$ and ending at a critical point at approximately ($\mu\simeq320$
MeV, $T\simeq75$ MeV ). This first-order line manifests itself in
Fig. \ref{fig:OrderParameters_constantT_A} in the discontinuous behavior
of $\left\langle {\rm Tr}_{F}P\right\rangle $, $\left\langle {\rm Tr}_{F}P^{\dagger}\right\rangle $
and $m$ at $T=50$ MeV.

\begin{figure}
\includegraphics[width=5in]{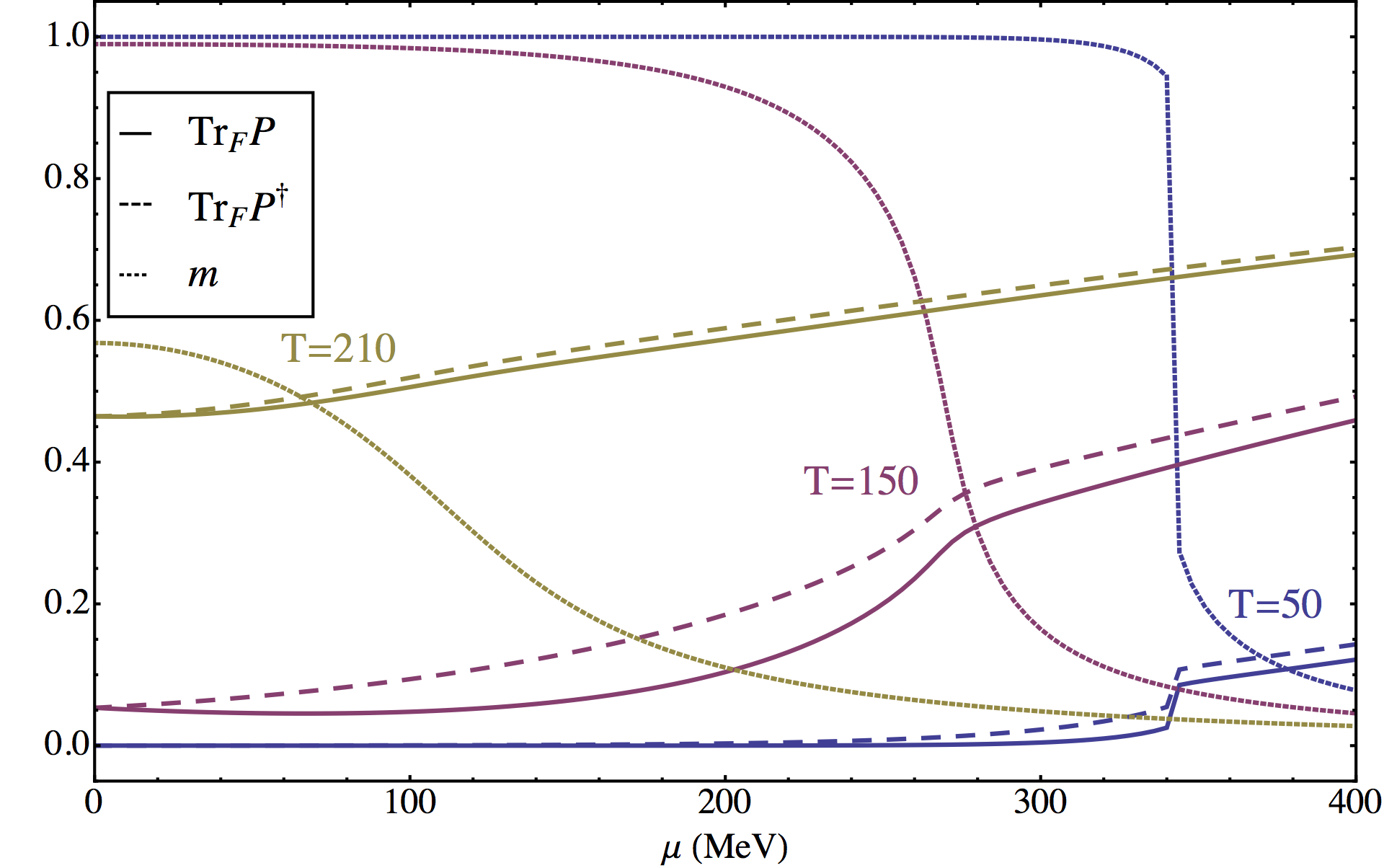}

\caption{\label{fig:OrderParameters_constantT_B}The constituent mass $m$,
$\left\langle {\rm Tr}_{F}P\right\rangle $ and $\left\langle {\rm Tr}_{F}P^{\dagger}\right\rangle $
as a function of $\mu$ for $T=50,\,150,\,\mbox{and 210\,\mbox{MeV}}$
for a PNJL model using Model B for confinement effects. The constituent
mass $m$ is normalized to one at $T=0$, and the Polyakov loops are
normalized to one in the limit as the temperature becomes large.}
\end{figure}

\begin{figure}
\includegraphics[width=5in]{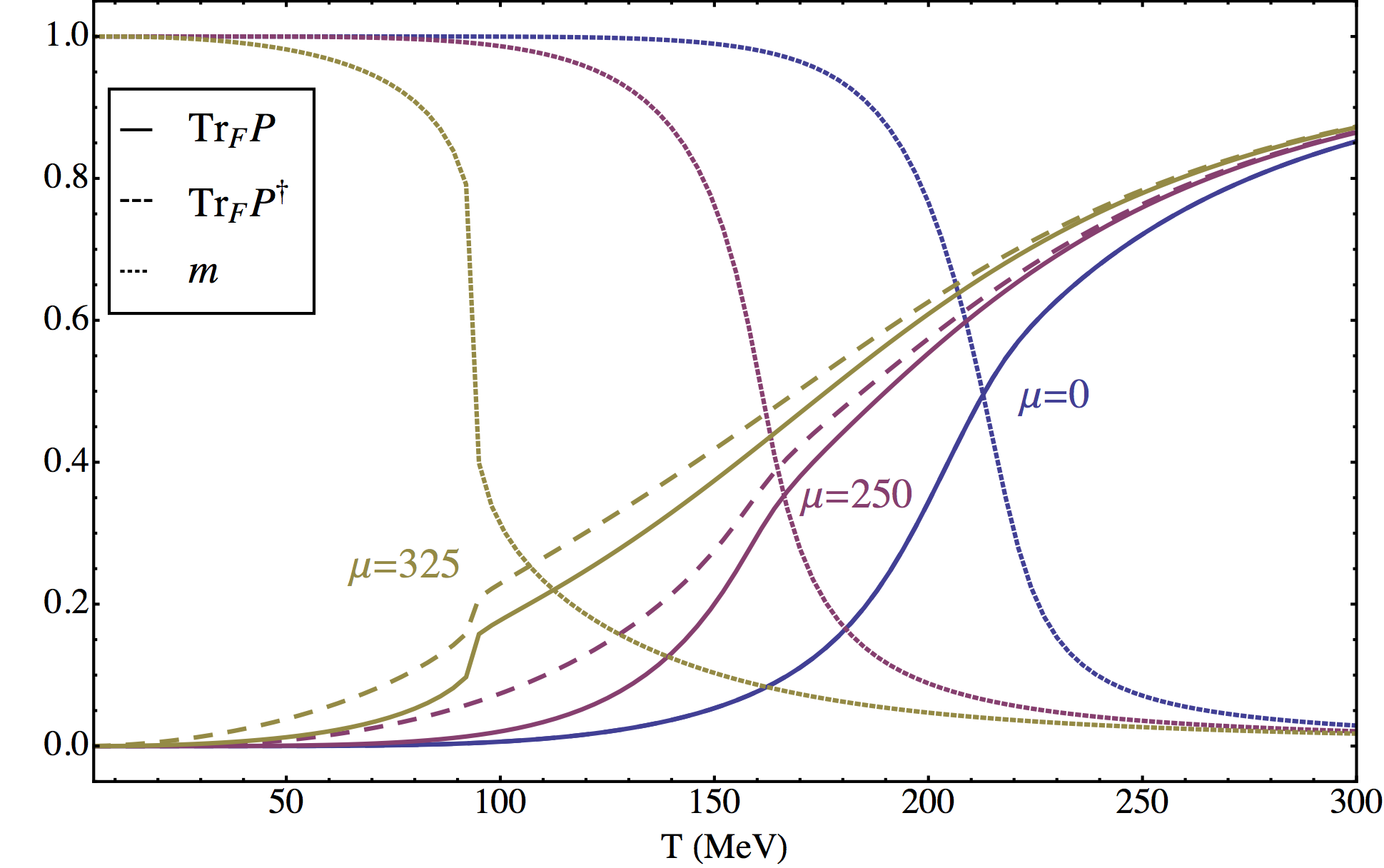}

\caption{\label{fig:OrderParameters_constantMu_B}The constituent mass $m$,
$\left\langle {\rm Tr}_{F}P\right\rangle $ and $\left\langle {\rm Tr}_{F}P^{\dagger}\right\rangle $
as a function of $T$ for $\mu=0,\,250$ and $325$ MeV for a PNJL
model using Model B for confinement effects. The constituent mass
$m$ is normalized to one at $T=0$, and the Polyakov loops are normalized
to one in the limit as the temperature becomes large.}
\end{figure}

Figures \ref{fig:OrderParameters_constantT_B} and \ref{fig:OrderParameters_constantMu_B}
show the corrsponding behavior of $\left\langle {\rm Tr}_{F}P\right\rangle ,\left\langle {\rm Tr}_{F}P^{\dagger}\right\rangle $
and $m$ using $V_{d}^{B}$ to implement confinement. In all figures
of this type, $m$ is normalized to its value at $(\mu=0,T=0)$, while
$\left\langle {\rm Tr}_{F}P\right\rangle $ and $\left\langle {\rm Tr}_{F}P^{\dagger}\right\rangle $
are normalized so that they go to one as $T$ goes to infinity. As
is Model A, only crossover behavior is seen at $\mu=0$. The critical
line starting at $\mu\approx350$ MeV when $T=0$ ends at a critical
point at approximately $(\mu=320\, MeV,\, T=100\, MeV)$. The first-order
line again manifests itself in Fig. \ref{fig:OrderParameters_constantT_B}
in the discontinuous behavior of $\left\langle {\rm Tr}_{F}P\right\rangle $,
$\left\langle {\rm Tr}_{F}P^{\dagger}\right\rangle $ and $m$ at
$T=50$ MeV.

\begin{figure}
\includegraphics[width=5in]{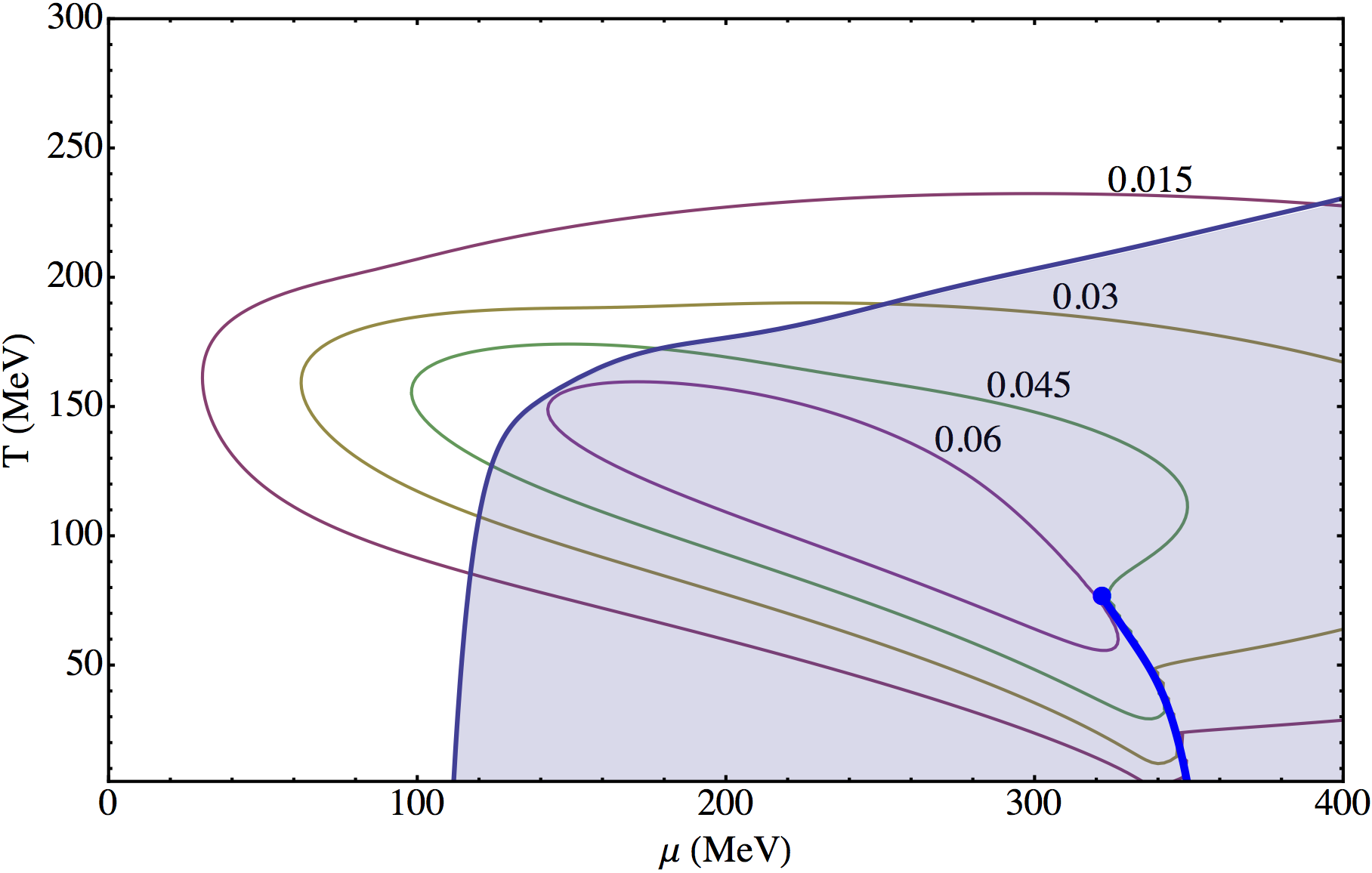}

\caption{\label{fig:PNJL_A_Psi_DisorderLine_CriticalLine}Contour plot of $\psi$
in the $\mu-T$ plane for a PNJL modle using Model A for confinement
effects. The region where $\kappa_{I}\ne0$  is shaded. The critical
line and its endpoint are also shown.}

\end{figure}

\begin{figure}
\includegraphics[width=5in]{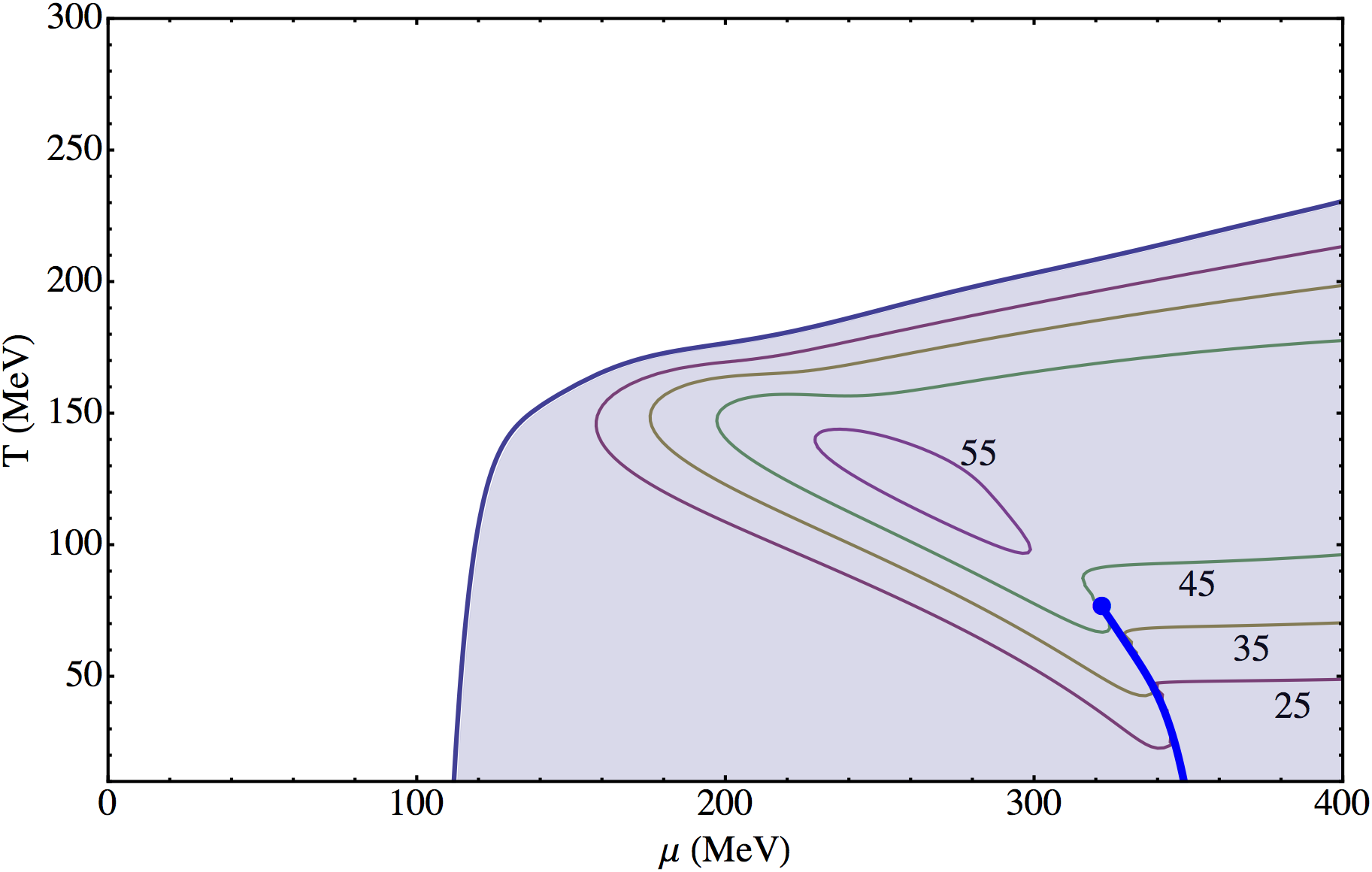}

\caption{\label{fig:PNJL_A_ImaginaryMass_DisorderLine_CriticalLine}Contour
plot of $\kappa_{I}$ in the $\mu-T$ plane for a PNJL modle using
Model A for confinement effects. Contours are given in MeV, with $\alpha_{s}$
set to one. The region where $\kappa_{I}\ne0$ is shaded. The critical
line and its endpoint are also shown.}
\end{figure}

Figure \ref{fig:PNJL_A_Psi_DisorderLine_CriticalLine} shows contour
lines for $\psi$ in the $\mu-T$ plane along with the region where
$\kappa_{I}\ne0$ as well as the critical line. The overall shape
of the disorder line is similar to that found in the previous section
for heavy quarks, but of course shifted to a much lower value $\mu$.
The critical line lies completely within the region $\kappa_{I}\ne0$.
Figure \ref{fig:PNJL_A_ImaginaryMass_DisorderLine_CriticalLine} shows
a contour plot for $\kappa_{I}$. In both figures, a jump in $\psi$
and $\kappa_{I}$ is visible as the critical line is crossed.

\begin{figure}
\includegraphics[width=5in]{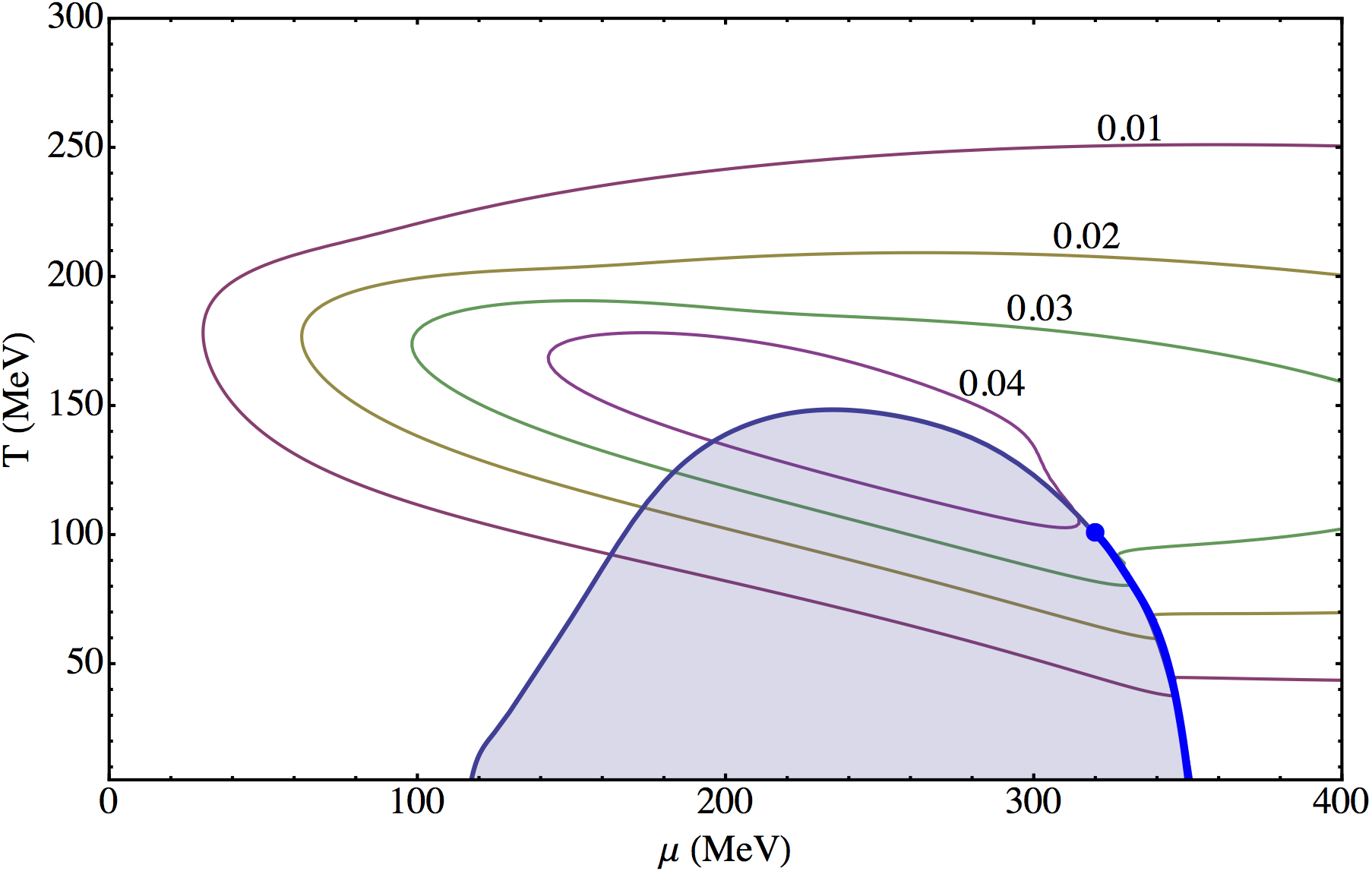}

\caption{\label{fig:PNJL_B_Psi_DisorderLine_CriticalLine}Contour plot of $\psi$
in the $\mu-T$ plane for a PNJL modle using Model B for confinement
effects. The region where $\kappa_{I}\ne0$ is shaded. The critical
line and its endpoint are also shown.}
\end{figure}

\begin{figure}
\includegraphics[width=5in]{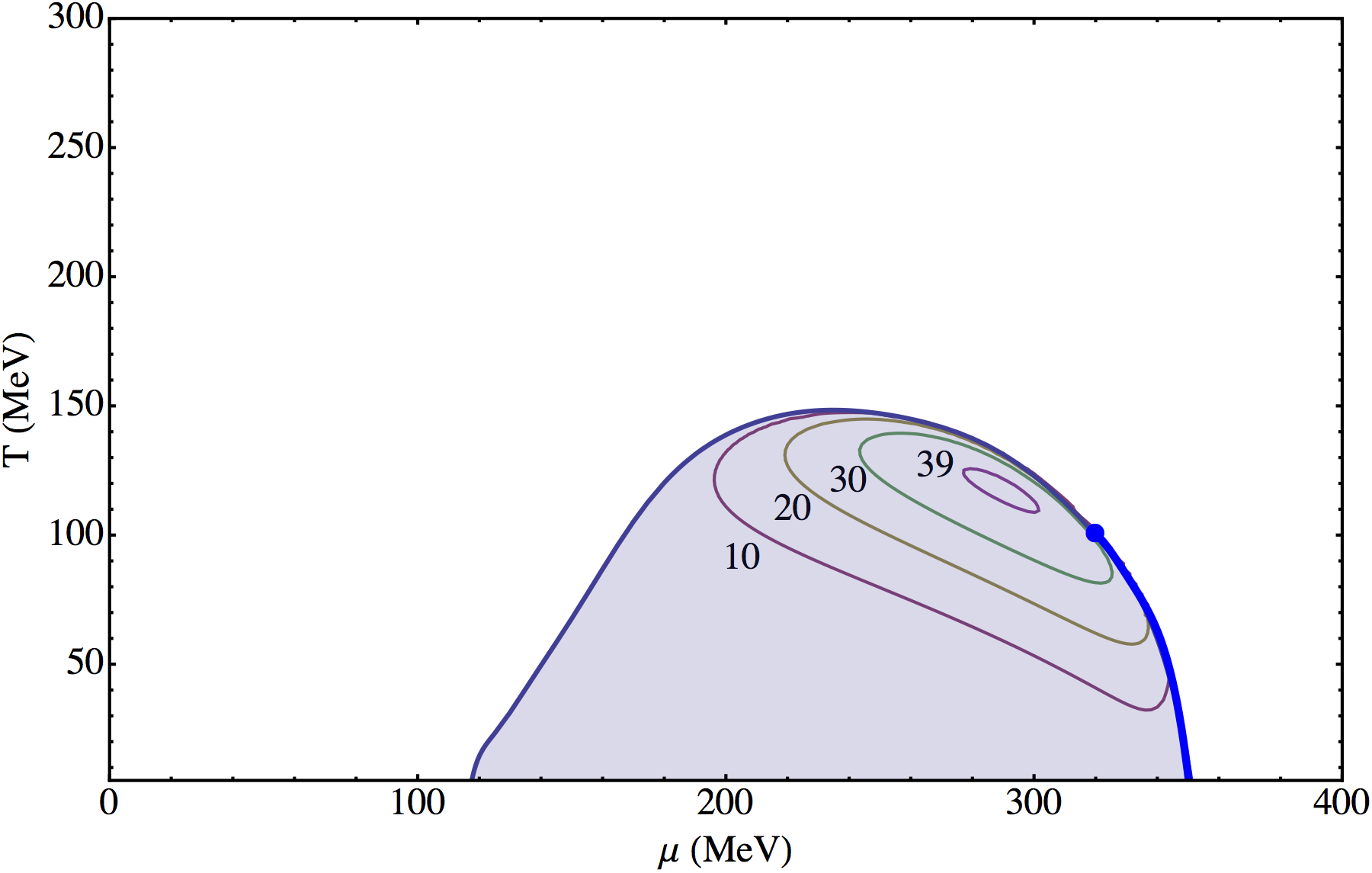}

\caption{\label{fig:PNJL_B_ImaginaryMass_DisorderLine_CriticalLine}Contour
plot of $\kappa_{I}$ in the $\mu-T$ plane for a PNJL modle using
Model B for confinement effects. Contours are given in MeV, with $\alpha_{s}$
set to one. The region where $\kappa_{I}\ne0$ is shaded. The critical
line and its endpoint are also shown.}
\end{figure}

As with Model A, the PNJL model using $V_{d}^{B}$ shows many of the
same features found for heavy quarks. Figure \ref{fig:PNJL_B_Psi_DisorderLine_CriticalLine}
contour lines for $\psi$ in the $\mu-T$ plane along with the region
where $\kappa_{I}\ne0$ as well as the critical line, and Figure \ref{fig:PNJL_B_ImaginaryMass_DisorderLine_CriticalLine}
shows a contour plot for $\kappa_{I}$. A striking difference between
Model A and Model B is that the critical line now lies on the boundary
of the region $\kappa_{I}\ne0$, and the disorder line appears to
be a smooth continuation of the critical line out of the critical
end-point.

\section{Conclusions}

As we have shown, the sign problem in QCD at finite density makes
it very desirable to extend real fields into the complex plane. This
extension is certainly necessary for steepest descents methods to
yield correct results. Complex saddle points lead naturally to $\left\langle TrP\right\rangle \ne\left\langle TrP^{\dagger}\right\rangle $,
a result that is much more difficult to obtain when fields are restricted
to the real axis. The nature of these saddle points are restricted
by $\mathcal{CK}$ symmetry. The case of a single dominant saddle
point is particularly tractable in theoretical analysis. In the class
of models we have examined, the saddle point is not far from the real
axis, as indicated by the small values of $\psi$ and corresponding
small differences between $\left\langle TrP\right\rangle $ and $\left\langle TrP^{\dagger}\right\rangle $.
This is good news for lattice simulation efforts, as it suggests only
a modest excursion into the complex plane is needed. The small value
of $\psi$ also indicates a small difference for thermodynamic quantities
such as pressure and internal energy between our work and previous
work on phenomenological models where only real fields were used.
For all six cases studied here, the maximum value of $\psi$ occurs
in the region where quark degrees of freedom are ``turning on,''
as indicated by crossover or critical behavior. In our previous work
on Model A for massless quarks \cite{Nishimura:2014rxa}, we were
able to show analytically how $\psi\ne0$ can arise from the interplay
of confinement and deconfinement when $\mu\ne0$, and our results
here are consistent. For the two PNJL models, it is striking that
the largest values of $\psi$ occur near the critical end point. These
predictions can be checked in lattice simulations by the direct measurement
of $\left\langle TrP\right\rangle $ and $\left\langle TrP^{\dagger}\right\rangle $
once sufficiently effective simulation algorithms are developed.

In all six cases studied, $\psi\ne0$ leads to two different eigenvalues
for the $A_{4}$ mass matrix. In five of the six cases studied, a
disorder line is found. This disorder line marks the boundary of the
region where the real parts of the mass matrix eigenvalues become
degenerate as the eigenvalues form a complex conjugate pair. In the
PNJL models, the disorder line is closely associated with the critical
line. Inside the region bounded by the disorder line, the complex
conjugate pairs gives rise to color charge density oscillations. Patel
has developed a scenario in which such oscillations might be observed
experimentally \cite{Patel:2011dp,Patel:2012vn}. Our results indicate
that the oscillations may have too large a wavelength to be directly
observable in experiment, although estimates based on phenomenological
models should be applied cautiously. The mass matrix eigenvalues are
in principle accessible in lattice simulations via the measurement
of Polyakov loop correlation functions. A direct determination of
$\kappa_{I}$ may be difficult, but the disorder line itself could
be determined from the merging of the values of $Re\left(\kappa_{1}\right)$
with $Re\left(\kappa_{2}\right)$. 

While the behavior of $\left\langle TrP\right\rangle $, $\left\langle TrP^{\dagger}\right\rangle $
and $\left\langle \bar{\psi}\psi\right\rangle $, as determined by
lattice simulations, do not strongly differentiate between the two
confining potential terms, Model A and Model B, the corresponding
two-point correlation functions do. The most physically relevant case
of PNJL models show both common features as well as clear differences
in the behavior of the disorder line between Model A and Model B.
In both cases, the maximum value of $\kappa_{I}$ occurs slightly
above and to the left of the critical end point in the $\mu-T$ plane,
in the vicinity of the region where the ratio $Tr_{F}P^{\dagger}/Tr_{F}P$
is largest. In Model A, the critical line is contained within the
boundary of the disorder line, but in Model B the disorder line appears
to come out of the critical end point as a continuation of the critical
line, a common behavior for disorder lines. Furthermore, in Model
A the disorder line continues diagonally in the $\mu-T$ plane for
large $\mu$ and $T$, but for Model B, the line bends over into the
critical line. With Model A there is thus a possibility that the effects
of the disorder line might be visible in the results of the Compressed
Baryonic Matter (CBM) experiment at FAIR. The disorder line also strongly
differentiates between Model A and Model B in the case of heavy quarks,
so lattice simulations of either light or heavy quarks that can locate
the disorder line have the potential to discriminate between the two
models.
\begin{acknowledgments}
We thank Zohar Nussinov for helpful discussions on disorder lines
in condensed matter physics.
\end{acknowledgments}
\bibliographystyle{unsrt}
\bibliography{Long_ms}

\end{document}